\documentclass[review,onefignum,onetabnum]{siamonline171218}

\usepackage{bm}
\usepackage{comment}
\usepackage{mathtools}
\usepackage{bbm}

\usepackage{lipsum}
\usepackage{amsfonts}
\usepackage{graphicx}
\usepackage{epstopdf}

\usepackage{algorithm} 
\usepackage{algpseudocode} 

\algnewcommand\algorithmicinput{\textbf{Input:}}
\algnewcommand\algorithmicoutput{\textbf{Output:}}
\algnewcommand\Input{\item[\algorithmicinput]}%
\algnewcommand\Output{\item[\algorithmicoutput]}%

\usepackage{tabularx}
\usepackage{booktabs}

\ifpdf
  \DeclareGraphicsExtensions{.eps,.pdf,.png,.jpg}
\else
  \DeclareGraphicsExtensions{.eps}
\fi

\usepackage{enumitem}
\setlist[enumerate]{leftmargin=.5in}
\setlist[itemize]{leftmargin=.5in}

\newsiamremark{remark}{Remark}
\newsiamremark{hypothesis}{Hypothesis}
\crefname{hypothesis}{Hypothesis}{Hypotheses}
\newsiamthm{claim}{Claim}

\newtheorem{experiment}{Experiment}

\DeclareMathOperator{\E}{\mathbb{E}}
\DeclareMathOperator{\V}{\mathbb{V}}

\DeclareMathOperator{\Cov}{Cov}

\DeclareMathOperator{\tr}{tr}
\DeclareMathOperator*{\argmax}{argmax} 

\DeclareMathOperator{\argmin}{argmin}

\usepackage{amsopn}

\ifpdf
\hypersetup{
  pdftitle={Fast data inversion for high-dimensional Ornstein-Uhlenbeck processes from noisy measurements},
  pdfauthor={Yizi Lin, Xubo Liu, Paul Segall and Mengyang Gu}
}
\fi

\headers{Fast data inversion}{Yizi Lin, Xubo Liu, Paul Segall and Mengyang Gu}

\title{Fast data inversion for high-dimensional {Ornstein-Uhlenbeck processes} from noisy measurements
}

\author{Yizi Lin, \thanks{Department of Statistics and Applied Probability, 
University of California, Santa Barbara, CA (\email{lin768@umail.ucsb.edu}). Equal contribution.}
\and Xubo Liu\thanks{Department of Statistics and Applied Probability, 
University of California, Santa Barbara, CA
  (\email{xubo@umail.ucsb.edu}). Equal contribution.}
\and Paul Segall\thanks{Department of Geophysics, Stanford University, CA
  (\email{segall@stanford.edu}).}
  \and Mengyang Gu\thanks{Department of Statistics and Applied Probability, 
University of California, Santa Barbara, CA
  (\email{mengyang@pstat.ucsb.edu}).}
}

\begin{document}

\maketitle

\begin{abstract}
In this work, we develop a scalable approach for a flexible latent factor model for high-dimensional dynamical systems. Each latent factor process has its own correlation and variance parameters, and the orthogonal factor loading matrix can be either fixed or estimated. We utilize an orthogonal factor loading matrix that avoids computing the inversion of the posterior covariance matrix at each time of the Kalman filter, and derive closed-form expressions in an expectation-maximization algorithm for parameter estimation, which substantially reduces the computational complexity without approximation. 
{ Our approach has several applications, including noise filtering for high-dimensional time series,  estimating nonseparable covariance structure between different time series, and estimating latent physical processes from real-world measurements.} Extensive simulated studies illustrate higher accuracy and scalability of our approach compared to alternatives. Furthermore,  by applying our method to geodetic measurements to estimate slow slip events from geodetic data in the Cascadia region, our estimated slip better agrees with independently measured seismic data of tremor events. The substantial acceleration from our method enables the use of massive noisy data for geological hazard quantification and other applications. 
\end{abstract}



\section{Introduction}
\label{section:intro}

Latent factor models of time-dependent systems have wide applications, such as estimating unobservable geophysical processes \cite{segall1997time}, {model calibration for computer simulation of multivariate outputs \cite{higdon2008computer}}, and inferring multiple time series in economics \cite{lam2012factor}. Though applications differ,  latent factor models can be broadly classified into two categories with either fixed or data-dependent factor loading matrices.

Basis functions, such as discrete Fourier basis and Green's functions, are widely used for inverse estimation of experimental or field observations. Differential dynamic microscopy \cite{cerbino2008differential}, for instance, is a physical approach for estimating the rheological  properties of the materials by microscopy videos. The estimation corresponds to minimizing the temporal autocorrelation in the Fourier space by a latent factor model with the complex conjugate of the discrete Fourier basis 
\cite{gu2024ab}. Green's functions, as another example, relate field observations to unobserved displacement at different spatial scales, such as cellular force estimation by traction force microscopy \cite{sabass2008high} and geologic slip estimation by geodetic data \cite{yabuki1992geodetic}. 

On the other hand, the factor loading matrix can be estimated by data. Probabilistic models were built to understand the underlying model assumptions made by these estimations. The loading matrix estimated by the principal component analysis \cite{berkooz1993proper}, for instance, 
is shown to be the maximum marginal likelihood estimator of a latent factor model with random factors independently distributed as standard Gaussian distributions 
\cite{tipping1999probabilistic}. 
 Various approaches extend the independent assumption for correlated data. 
 In \cite{gu2024probabilistic}, for instance, the authors show the estimation of the dynamic mode decomposition  is equivalent to the maximum likelihood estimator of the linear mapping matrix in a vector autoregressive model of noise-free observations. 
 As another example,   each latent factor is modeled by a Gaussian process in   coregionalization models \cite{gelfand2004nonstationary}. 

In this work, we develop a scalable and efficient approach for latent factor models with correlated factors and an orthogonal factor loading matrix, either fixed or estimated by data.  Our contributions are threefold. First, estimating a large number of correlation parameters in multivariate Gaussian processes is a fundamental challenge. 
When each latent factor process has distinct parameters, conventional strategies, such as posterior sampling or numerical optimization   
can be prohibitively slow for estimating a large number of parameters. \cite{gu2024probabilistic}, which can be both costly. 
We developed a novel expectation-maximization (EM) algorithm for fast parameter estimation, {which avoids expensive matrix inversion at each time in the Kalman filter required in the conventional EM algorithm for state space models \cite{shumway1982approach},  based on the orthogonal projection through the latent factor loading matrix}. 
We surprisingly found that the estimation of the factor loading matrix, correlation, and variance parameters all have a closed-form expression in the algorithm. 
In particular, the correlation parameters of latent processes can be solved by cubic equations of order three. {These key results are provided in Theorem \ref{theorem:closed_form_expression_em}.   The log likelihood in the EM algorithm typically converges in less than 20 iterations in most of the scenarios studied in this work, an example of which is shown in Figure \ref{fig:diffusion}. The fast convergence makes our fast approach an almost exact solution for many problems.} 
Second, we model latent factors by a multivariate Ornstein-Uhlenbeck process that contains distinct correlation and variance parameters, which provide a flexible way to capture dependence across output coordinates. 
Third, we develop a new way to estimate the number of factors with either a fixed or an estimated factor loading matrix by matching the estimated noise variance with its measurement. 

{Our approach has advantages over alternative approaches developed in computational mathematics, statistics, and geophysics communities. 
In computational mathematics, dynamic mode decomposition is a popular approach that linearizes the one-step-ahead transition operator of nonlinear dynamical systems to reconstruct the dynamics by the eigenpairs of the
linear mapping matrix \cite{schmid2010dynamic,tu2014dmd}, which produces a finite-dimensional approximation of the Koopman modes and eigenvalues \cite{mezic2013analysis}. Our model extends the probabilistic model of the dynamic mode decomposition by including the noise model and utilizing a symmetric factor loading matrix for reducing the space in estimation. Significant improvements against dynamic mode decomposition will be shown for noisy observations in Section \ref{sec:simulations}. 
}

{In the statistics community, EM algorithm and Kalman filter have long been used for estimating parameters in the vector autoregressive models \cite{shumway1982approach,metaxoglou2007maximum}, and they are implemented in 
\cite{astsa}. Our approach contains three advantages over the conventional EM algorithm for autoregressive models. First, our approach enables both the estimated factor loading matrix and dimension reduction of latent factors, while the EM algorithm typically assumes that the dimension of the latent factor is the same as the number of time series. Second, assuming models with $d$ latent factor processes at $n$ time points, our new algorithm only requires $\mathcal O(nd)$ operations in Kalman filter without approximation due to the orthogonality of the latent factor loading matrix, while the conventional EM algorithm for vector autoregressive models requires $\mathcal O(nk^3)$ operations for inverting a $k\times k$ one-step-ahead predictive covariance matrix of the observations at each time step in Kalman filter.  Third, our model can be set to be either stationary or non-stationary, yet the estimation from the conventional EM algorithm in \cite{shumway1982approach} cannot guarantee the stationarity. 
}

In the geophysics community, the network inverse filter \cite{segall1997time,bartlow2011space} is widely used for estimating geologic slips from ground deformation measurements, such as the GPS time series. It assumes the vector autoregressive processes with the latent states following integrated Brownian motions, with shared parameters across different processes. Our method is both faster and more flexible with distinct correlation and variance parameters for each latent process, as will be shown in Section \ref{sec:simulations} and Section \ref{sec:real_application}.

The rest of the paper is organized as follows. In Section \ref{sec:fmou}, we introduce {a motivating example for estimating geologic slip} and introduce a flexible latent factor model for high-dimensional dynamical systems with a fast EM algorithm given in Theorem  \ref{theorem:closed_form_expression_em}. 
We show our model substantially accelerates the computation compared to other approaches in Section \ref{sec:connections}. Extensive experiments in Section \ref{sec:simulations} show high accuracy and computational scalability of our approach. In Section \ref{sec:real_application}, we compare our approach with existing ways to estimate the slip migration in the Cascadia region using the GPS data and demonstrate the higher detection rate of tremor events from an independent source of seismic data not used in estimation. We conclude our study and outline future directions in Section \ref{sec:conclusion}. {The FMOU algorithm is coded in the {\sf FastGaSP 0.6.2} package available on CRAN \cite{FastGaSP}. The data and code are made publicly available in GitHub: \url{https://github.com/UncertaintyQuantification/FMOU/}.}

\section{Fast and efficient estimation of high-dimensional dynamical systems}
\label{sec:fmou}
\subsection{Motivating example:  Geologic slip estimation by the  ground deformation data}
Our study is motivated by geologic slip estimation from geodetic measurements, but the new approach has broad applications for problems in science and engineering. Ground deformation observations, such as continuous Global Positioning System (GPS) observations \cite{segall1997time} and interferometric synthetic-aperture radar interferograms \cite{anderson2019magma}, have been widely used to quantify geological hazards, including volcanic eruptions and earthquakes. The goal of our application is to estimate  slip that quantifies the relative movement velocities of two sides of the faults \cite{burgmann2018geophysics}. Since slip cannot be directly observed,  time-dependent ground deformation information from GPS data has been used for slip estimation \cite{bock1993detection}.  

The Cascadia subduction zone is known to experience aseismic, transient slip events known as Slow Slip Events  \cite{gomberg2010slow}. Slow slip events, recorded by high precision GPS data, have been found to be spatially and temporally associated with `tectonic tremor', composed of low-frequency earthquakes recorded on seismic stations. We utilized the GPS data in the Cascadia subduction zone, publicly available at the Plate Boundary Observatory, for estimating the slip on the megathrust fault from June 2011 to August 2011 \cite{bartlow2011space}. 
Denote the $p_y$-dimensional observations $\mathbf y(\mathbf x, t) \in \mathbb R^{p_y}$ at location $\mathbf x \in \mathbb R^{p_x}$ and time $t \in \mathbb R$. In our application, as the vertical displacement measurements contain little information for  slip migration, 
the ground displacement measurement $\mathbf{y}(\mathbf x, t)$ is a two-dimensional vector in the East-West and North-South directions observed at equally spaced time $t$ at a GPS station, with spatial coordinates $\mathbf x=(x_1, x_2)$, and hence $p_x=p_y=2$. 
The network inversion filter  introduced in 
\cite{segall1997time} is a popular approach for modeling the displacement at the Earth's surface by: 
$\mathbf{y}(\mathbf x, t)=\int \mathbf{G}(\mathbf x, \bm \xi) \mathbf z_s(\bm \xi, t) d\bm \xi + \bm{\mu}(\mathbf x, t) + \bm{\epsilon}(t), $
where  $\mathbf{G}(\mathbf x,\bm \xi)$  is a quasi-static elastic Green's function \cite{aki1980quantitative} that relates the observations to the unobservable slip $\mathbf z_s(\bm \xi, t)$, and  $\bm{\mu}(\mathbf x, t)$, the local benchmark motion, captures the site-specific local  motion of the GPS antenna. 
For data measured over a short period, such as the transient deformations from 2011 in the Cascadia subduction zone considered in this work, the impact of local motion is negligible, and hence we let $\bm{\mu}(\mathbf x, t) =\mathbf 0$. 
Furthermore, {{the reference frame motion is often integrated into this equation by augmented latent processes}}.  The Gaussian noise of the measurement is denoted by $\bm{\epsilon}(t)$ and the  variance  of the noise is typically available {\it a priori} from the GPS measurements.

Suppose we collect GPS observations at $\tilde k$ locations, resulting in a $k= p_y\tilde k$ vector of observations $\mathbf y(t)=[\mathbf y(\mathbf x_1,t),\mathbf y(\mathbf x_2, t),\dots, \mathbf y(\mathbf x_{\tilde{k}},t)]^T$. In \cite{segall1997time}, the output vector is modeled as:  
\begin{equation}
	\mathbf y(t) = {\mathbf{G}}
    \mathbf z_s(t)+\bm \epsilon (t),
 \label{equ:displacement}
\end{equation}
where $\mathbf G$ is a $k\times k'$ matrix of discretized Green's function,  $ \mathbf z_s(t) =[z_s(\bm \xi_1, t), \dots,  z_s(\bm \xi_{k^{\prime}},t)]^T$  is a $k'$-vector of unobservable geologic slip,  with subscript `s' meaning the slip, and the Gaussian noises follow $\bm \epsilon(t) \sim \mathcal{MN}(\mathbf 0, 
\sigma^2_0 \mathbf I_k)$ with variance $\sigma^2_0$. 
The $(i,h)$ block of $\mathbf G$ is $\mathbf G_{i,h}=\mathbf{G}(\mathbf x_i, \bm \xi_h)\Delta$ and $\Delta$ is the area size in discretization,  for $i=1,...,\tilde k$ and $h=1,..., k'$ 
with $\tilde k=100$ GPS stations and $k'=1978$ discretization points used in estimating the slip propagation in Cascadia \cite{bartlow2011space}. 
We found that increasing the number of discretization points  
has almost no impact on estimates of the slip. 
 
Let $\mathbf U_0$ be $k\times d$ left singular vector matrix corresponding to the largest $d$ singular values in the singular value decomposition (SVD) of the Green's function $\mathbf G \approx \mathbf U_0  \mathbf D_0 \mathbf V_0^T$, where $\mathbf D_0 $ is a $d\times d$ diagonal matrix and $\mathbf V_0$ is a $k'\times d$ matrix of $d$ right singular vectors with $d\leq k$. 
 {As the number of discretization points $k'$ is larger than $k$,   the latent slip vector is modeled by a linear combination of the latent processes \cite{segall1997time} to obtain a reduced order representation}: 
\begin{equation}
    \mathbf{z}_s(t) = {\mathbf G}^T\mathbf U_0 \mathbf {\tilde z}(t),
    \label{equ:reconstructed_slip}
\end{equation}  
with $\mathbf {\tilde z}(t)$ being a $d$-vector of latent processes. {{The latent  process $\tilde z_l(t)$ is assumed to follow an integrated Brownian motion  in \cite{segall1997time}.}} The Kalman filter is used for computing the posterior distribution of the latent process $\mathbf {\tilde z}(\cdot)$ and slip estimation. There are several restrictions in this approach. First, the latent process ${\tilde z}_l(\cdot)$ has the same covariance across  $l=1,...,d$, yet the random latent factors can have different correlation length scales. Second, one needs to invert a $k\times k$ covariance matrix at each time in the Kalman filter, which is prohibitively slow when the number of GPS stations is moderately large. Third, selecting the number of latent factors is not discussed in this approach. 

The goal of this work is to develop a generally applicable approach and scalable algorithm for latent models for estimating the mean of the data and the posterior distribution of the latent variables. For estimating the slip rates in Cascadia between June and August in 2011, we found that the estimated slips had higher spatial correlation with independently detected tremor events not used in estimation. Furthermore, our method is {$500-1600$} times faster than the {network inversion filter in \cite{segall1997time}} in our real application. The high scalability of our method enables abundant data from GPS networks to be jointly used for hazard quantification. 

\subsection{A multivariate Ornstein-Uhlenbeck process with an orthogonal basis matrix} 
\label{subsec:MOU}
We follow the assumption in (\ref{equ:reconstructed_slip}), which indicates that the mean in   (\ref{equ:displacement}) is approximated by $ {\mathbf{G}}\mathbf z_s(t)\approx \mathbf U_0 \mathbf z(t)$, where $\mathbf z(t)=\mathbf D^2_0 \mathbf {\tilde z}(t)$ with the $l$th entry being $z_l(t)$, and there is no approximation when $k=d$. The $d$-dimensional latent factor processes $\mathbf z(\cdot)$ are modeled by independent Ornstein-Uhlenbeck (OU) processes, where each process has its own correlation and variance parameters. {The OU processes correspond to autoregressive models of order 1, widely used for modeling time series. Furthermore, the proposed model is closely related to the dynamic mode decomposition, a popular dimension reduction tool to approximate dynamical systems \cite{schmid2010dynamic}, as will be compared in Section \ref{sec:connection_DMD}.}
Together, Equations (\ref{equ:displacement})-(\ref{equ:reconstructed_slip}) lead  to 
\begin{align}
    \mathbf y(t) &=  \mathbf U_0 {\mathbf z}(t) +  \bm \epsilon(t), \label{equ:FMOU} \\
    {z}_l(t) &= \rho_l {z}_l(t-1) + w_l(t), \label{equ:OU}
\end{align}
where $\mathbf U_0$ is a $k \times d$ orthogonal factor loading matrix, either fixed or estimated, which relates the $d$-dimensional latent factors to $k$-dimensional measurements with $d\leq k$. 
Here, the parameter $\rho_l$ controls the temporal correlation length of the $l$th latent process, and $w_l(t)\sim \mathcal{N}(0, \sigma^2_l)$ is the innovation of the $l$th latent process for $t=2,...,n$, with the initial state being $z_l(1)\sim \mathcal N(0,\tau^2_l)$ for $l=1,\dots, d$.  {The $d$ sets of parameters $\bm{\rho}=(\rho_1,\rho_2,\dots, \rho_d)$ and $\bm{\sigma}^2=(\sigma_1^2,\sigma_2^2, \dots, \sigma_d^2)$ make the model flexible}. We postpone the discussion of estimating $d$ in Section \ref{subsec:dim_selection}.  

{ We highlight that  the orthogonal factor loading matrix $\mathbf U_0$  in Equation (\ref{equ:FMOU}) and OU processes in Equation (\ref{equ:OU}) bring substantial computational advantages, as will be elaborated soon.  The assumption of orthogonality of  $\mathbf U_0$ originates from several previous approaches. First, the matrix $\mathbf G$ in Equation (\ref{equ:displacement}) is typically neither symmetric nor orthogonal, but with the assumption in Equation (\ref{equ:reconstructed_slip}) used in \cite{segall1997time}, the model in (\ref{equ:displacement}) reduces to the Model (\ref{equ:FMOU}) with an orthogonal $\mathbf U_0$ matrix. Second, due to the non-identifiability between factor loadings and factors, the orthogonal assumption of factor loadings was also used in the past \cite{lam2011estimation}. In our experience, when the dimension $k$ is large, the orthogonal assumption typically works well, as the orthogonal matrix $\mathbf U_0$ has a large number of parameters that can be estimated, making the model flexible to capture dependence across different factor processes.} In this work, we develop novel algorithms for both fixed and estimated $\mathbf U_0$, which are broadly applicable to many applications. Furthermore, we assume {the mean of the factor processes to be zero for simplicity}, though additional mean structures can be included in the model. 

Lemma \ref{lemma:discrete_MOU}  connects the model in (\ref{equ:FMOU})-(\ref{equ:OU}) to the continuous-time multivariate Ornstein-Uhlenbeck process \cite{gardiner1985handbook, meucci2009review} with an orthogonal basis matrix. The derivations of all lemmas in this subsection are provided in Section SM1
in supplementary materials.  

\begin{lemma}
    \label{lemma:discrete_MOU}
    Denote the mean of the $k$-dimensional processes in (\ref{equ:FMOU}) by $\mathbf m(t)=\mathbf U_0 \mathbf z(t)$.  When {$\rho_l \in (0,1)$ in Equation (\ref{equ:OU}) for $l=1,\dots,d$}, the mean process $\mathbf m(t)$  is a discretized process of the continuous-time multivariate Ornstein-Uhlenbeck process, defined by the  stochastic differential equation:
    \begin{equation}
        \label{equ:continuous_MOU}
            d \mathbf m(t) = -\mathbf U_0 \mathbf{D} 
                \mathbf U^T_0 \mathbf m(t)dt + \mathbf U_0 \mathbf{\tilde D} d\mathbf{B}_t,
    \end{equation}
   where $\mathbf{D}=\text{diag}(-\log(\rho_1),\dots, -\log(\rho_d))$, 
   $\tilde{\mathbf D}$ is a diagonal matrix with the $l$th element being $\sqrt{\frac{-2\sigma_l^2 \log(\rho_l)}{1-\rho_l^2}}$ for $l=1,\dots, d$, and $\mathbf B_t$ is a vector of independent Brownian motions. 
\end{lemma}

We study the model in (\ref{equ:FMOU})-(\ref{equ:OU}) because of its flexibility due to distinct parameters $(\rho_l, \sigma_l^2)$ for each latent factor $l$, for $l=1,...,d$ and the computational scalability when the number of latent factors, $d$, is large. The latent process $z_l(\cdot)$ can be either stationary or nonstationary, depending on the assumption of the initial state, stated in the Lemma \ref{lemma:OU_stationary_conditions}. 
\begin{lemma}
For $l=1,...,d$, the covariance of the  process $z_l(\cdot)$ in  Equation (\ref{equ:OU}) follows 
\begin{equation}
\label{equ:cov_z_general}
\mbox{Cov}[z_l(t),z_l(t')]= \rho_l^{t'-t}\left\{\rho_l^{2(t-1)}\tau_l^2 + \frac{1-\rho_l^{2(t-1)}}{(1-\rho_l^2)}\sigma_l^2\right\}, \quad \text{for } t'\geq t.
\end{equation}
In particular, when the variance of initial state is $\tau^2_l=\frac{\sigma^2_l}{1-\rho^2_l}$,  $z_l(\cdot)$ is stationary with  
\begin{equation}
\mbox{Cov}[z_l(t),z_l(t')]=\frac{\sigma^2_l}{1-\rho^2_l} \rho_l^{|t-t'|}.  
\label{equ:stationary_OU}
\end{equation}
\label{lemma:OU_stationary_conditions}
\end{lemma}

{When $\tau^2_l=\frac{\sigma^2_l}{1-\rho^2_l}$,  the latent  process $z_l(\cdot)$ is constrained to be stationary in Lemma \ref{lemma:OU_stationary_conditions}, and when $\tau^2_l$ is  estimated as a separate parameter in the EM algorithm, the latent factor process  $z_l(\cdot)$  may not be stationary. } 
In the rest of the article, we assume that each latent process is stationary by letting $\tau^2_l=\frac{\sigma^2_l}{1-\rho^2_l}$. 
Lemma \ref{lemma:R_inv_det} outlines key properties of the covariance and precision matrices, which enable us to modify Kalman filter (KF) \cite{kalman1960new} and Rauch–Tung–Striebel (RTS) smoother \cite{rauch1965maximum} to scalably compute expensive quantities in our fast algorithm. 

\begin{lemma}
\label{lemma:R_inv_det}
    Assume $\tau^2_l=\frac{\sigma^2_l}{1-\rho^2_l}$ and denote $\mathbf{\Sigma}_l=\frac{\sigma_l^2}{1-\rho_l^2}\mathbf{R}_l$ as the covariance matrix of $\mathbf z_l=(z_l(1),\dots, z_l(n))^T$ defined in  Equation (\ref{equ:OU}). The $(t,t')$ entry of $\mathbf R_l$ is $\rho_l^{|t-t'|}$, for $t,t'=1,2,\dots, n$. The inverse of $\mathbf R_l$ has a tri-diagonal structure, with diagonal entries of $\frac{1}{1-\rho_l^2}$ at the first and last positions, and $\frac{1+\rho_l^2}{1-\rho_l^2}$ at the remaining positions. The primary off-diagonal entries are $-\frac{\rho_l}{1-\rho_l^2}$. Additionally, the determinant of 
     $\bm{\Sigma}_l$ follows $|\bm{\Sigma}_l|=\frac{\sigma_l^{2n}}{1-\rho_l^2}$.
\end{lemma}

\subsection{A fast EM algorithm with closed-form expressions}
\label{subsec:EM_algorithm}

In this section, we derive a fast EM algorithm, named a Fast algorithm of Multivariate Ornstein-Uhlenbeck processes (FMOU), for a general scenario where the parameters are $\mathbf{\Theta}=(\mathbf U_0, \sigma_0^2, \bm{\rho}, \bm{\sigma}^2)$ with $\bm{\rho}=(\rho_1,\rho_2,\dots, \rho_d)$ and $\bm{\sigma}^2=(\sigma_1^2,\sigma_2^2, \dots, \sigma_d^2)$. The algorithm leverages KF \cite{kalman1960new} and RTS smoother \cite{rauch1965maximum} for computing required quantities with orthogonal projection, to bypass costly matrix inversion,  conventionally required in each step of the KF. A short review of the KF and RTS smoother  \cite{West1997,petris2009dynamic} is provided in Section SM2.   
 The proofs in this subsection are given in Section SM3  
in the supplementary materials.

Let $\mathbf Y = [\mathbf y(1),\dots, \mathbf y(n)]$ represent a $k \times n$ observation matrix and 
$\mathbf Z = [\mathbf z_1,\mathbf z_2, \cdots, \mathbf z_d]^T$ denotes a $d \times n$ latent factor matrix.
After integrating out the latent factors,  the parameters $\bm \Theta$ are estimated by the maximum marginal likelihood estimator (MMLE): 
\begin{align}
        {\bm \Theta}^{\text{MMLE}} & = \argmax_{\bm \Theta}  \int p(\mathbf{Y} \mid \mathbf Z, \bm \Theta) p(\mathbf Z \mid \bm \Theta) d \mathbf Z.
        \label{equ:mmle}
\end{align}
Direct optimization of the marginal likelihood of the model in Equations (\ref{equ:FMOU}) and (\ref{equ:OU}) 
can be unstable due to optimizing the latent factor matrix in the Stiefel manifold in each step of the optimization \cite{wen2013feasible} and numerical optimization in high-dimensional parameter space. We develop an EM algorithm which has closed-form expressions in each iteration. 
 
First, {we show in Section SM 3.1 in the Supplementary Materials that} the natural logarithm of the joint likelihood of $(\mathbf Y, \mathbf Z \mid \bm{\Theta})$ follows
\begin{align}
    &\ell(\bm \Theta)
    =C -\frac{n k}{2}\log(\sigma^2_0)- \frac{\tr(\mathbf Y^T \mathbf Y - 2\mathbf Y^T \mathbf U_0 \mathbf Z)}{2\sigma^2_0}-\sum^{d}_{l=1} \Bigl(\frac{\mathbf z_l^T \mathbf z_l}{2\sigma^2_0}  + \frac{ \log|\bm \Sigma_l |+\mathbf z_l^T \bm \Sigma^{-1}_l \mathbf z_l}{2}\Bigr),
 \label{equ:join_log_lik}
\end{align}
where $C=-\frac{(nk+nd)}{2}\log(2\pi)$ is a constant. 

In the E step, we calculate the expectation of the joint log-likelihood function with respect to the distribution of $\mathbf Z$ conditional on observations $\mathbf Y$, and the current estimate of parameters, denoted as $\bm {\hat \Theta}=(\mathbf{\hat U}_0, \hat{\sigma}_0^2, \bm{ \hat \rho}, \bm{ \hat \sigma}^2)$. { Our model in (\ref{equ:FMOU})-(\ref{equ:OU}) is 
  a specific case of the model in \cite{gu2020generalized} with the latent factor processes assumed to be OU processes, and this specification enables our model to  be computed much faster, as will be compared in Section \ref{sec:connection_gppca}. 
The posterior distribution of factors given $\bm{\hat \Theta}$
follows a multivariate Gaussian distribution shown in \cite{gu2020generalized}}: 
\begin{align}
    (\mathbf z_l \mid \mathbf{Y}, \bm{\hat \Theta}) \sim \mathcal{MN} \left(  \mathbf{\hat{z}}_l, \hat{\sigma}^2_0 \bm{\hat \Sigma}_l(  {\bm{\hat \Sigma}}_l+ \hat{\sigma}^2_0\mathbf I_n)^{-1} \right),
\label{equ:posterior_tilde_zl}
\end{align}
where $\mathbf{\hat{z}}_l =\bm{ \hat \Sigma}_l (\bm{\hat \Sigma}_l + \hat{\sigma}^2_0 \mathbf I_n)^{-1} \tilde {\mathbf y}_l$ with $\tilde{\mathbf Y} = \mathbf U_0^T \mathbf Y = [\tilde{\mathbf y}_1,\cdots, \tilde{\mathbf y}_d]^T$ being the projected observation matrix, and {the matrix $\hat{\mathbf \Sigma}_l$ is formed by the same kernel function as $\mathbf{\Sigma}_l$, with the current estimate of the parameters $\sigma^2_l$ and $\rho_l$ plugged in},  for $l=1,\ldots,d$. Denote $\mathbf{\hat Z}=[\mathbf{\hat z}_1,\dots,\mathbf{\hat z}_d]^T$, a $d\times n$ matrix of the posterior mean of latent factors, and $\bar \ell(\bm \Theta) = \mathbb{E}_{\mathbf Z \mid \mathbf{Y}, \bm{\hat \Theta} }[ \ell(\bm \Theta)]$. Utilizing  Equation (\ref{equ:posterior_tilde_zl}),  { we show in Section SM 3.2 in the Supplementary Materials that} the E step follows  
\begin{align}
\bar \ell(\bm \Theta) &= C + \frac{\mbox{tr}(\mathbf{Y}^T \mathbf U_0 \mathbf{\hat{\mathbf Z}})}{ \sigma^2_0}  - \sum^{d}_{l=1} \left\{ \frac{\log|\bm \Sigma_l |}{2} + \frac{ \mathbf{\hat z}_l^T \mathbf{\hat {z}}_l+\mbox{tr}[{\hat \sigma}_0^2 \bm{ \hat \Sigma}_l(  \bm{\hat \Sigma}_l+{\hat \sigma}_0^2\mathbf I_n)^{-1}]  }{2 \sigma^2_0} \right\} \nonumber \\
&\quad - \sum^{d}_{l=1} \frac{ \mathbf{\hat z}_l^T \bm \Sigma^{-1}_l \mathbf{\hat z}_l+\mbox{tr}[{\hat{\sigma}}_0^2 \bm \Sigma^{-1}_l \bm{ \hat \Sigma}_l(  \bm{\hat \Sigma}_l+{\hat{\sigma}}_0^2\mathbf I_n)^{-1}] }{2} -\frac{n k}{2}\log(\sigma^2_0)- \frac{\tr(\mathbf Y^T \mathbf Y)}{2\sigma^2_0},
\label{equ:e_step}
\end{align} 
where $\bm \Sigma_l$ depends on unknown $(\bm{ \rho},  \bm{\sigma}^2)$  to be optimized, and $\bm{\hat \Sigma}_l$ is obtained based on the current estimate of the parameters  $(\bm{ \hat \rho}, \hat{\bm{\sigma}}^2)$. 

Directly computing the conditional distribution in  Equation (\ref{equ:posterior_tilde_zl}) requires $\mathcal{O}(d n^3)$ operations due to inverting $d$  matrices of size $n\times n$, which can be computationally expensive. Instead, by treating the projected observations $\tilde{\mathbf y}_l=(\tilde y_l(1),\dots, \tilde y_l(n))^T$ as the noisy observations of the $l$th latent process, we can apply KF and RTS smoother independently to each latent factor. This approach efficiently computes the required quantities with  $\mathcal{O}(dn)$ operations, avoiding the need for matrix inversion. Lemma \ref{lemma:fast_comp_KF}  demonstrates their roles in simplifying computations in the E step. {The proof can be found in Supplementary materials SM3.3.}
\begin{lemma}
\label{lemma:fast_comp_KF}
Consider a dynamic linear model $\tilde y_l(t)=z_l(t)+\epsilon$ with latent factor process defined in (\ref{equ:OU}) and $\epsilon$ being an independent noise with variance $\sigma^2_0$, for $l=1,...,d$. Denote  
$s_l(t)= \E[z_l(t) \mid \mathbf {\tilde y}_l, \hat{\bm{\Theta}}]$, 
$S_l(t)= \V[z_l(t) \mid \mathbf {\tilde y}_l, \hat{\bm{\Theta}}]$ and $\tilde{S}_l(t) = \Cov[z_l(t), z_l(t+1) \mid \mathbf {\tilde y}_l, \hat{\bm{\Theta}}]$, which are computed by the KF and RTS smoother. 
We have
\begin{align}
     &\tr[{\hat \sigma}_0^2 \bm{ \hat \Sigma}_l(   \bm{\hat \Sigma}_l+{\hat \sigma}_0^2\mathbf I_n)^{-1}]=\sum_{t=1}^n\V[ z_l(t) \mid \mathbf Y,\hat{\bm \Theta}] = \sum_{t=1}^n S_l(t),\\
    & \mathbf{\hat z}_l^T \mathbf{\Sigma}_l^{-1} \mathbf{\hat z}_l = 
   \frac{1}{\sigma_l^2}\left((1-\rho_l^2){s_l^2(1)} + \sum_{t=2}^n(s_l(t)-\rho_ls_l(t-1))^2 \right),\\
   & \tr[{\hat{\sigma}}_0^2 \bm \Sigma^{-1}_l \bm{ \hat \Sigma}_l(  \bm{\hat \Sigma}_l+{\hat{\sigma}}_0^2\mathbf I_n)^{-1}] = \frac{1}{\sigma_l^2}\left(\sum_{t=1}^nS_l(t)+\rho_l^2\sum_{t=2}^{n-1}S_l(t)- 2\rho_l\sum_{t=1}^{n-1}\tilde{S}_l(t) \right).
\end{align}
\end{lemma}
In the M step, the parameters $\bm{\hat \Theta}^{\text{new}}	=(\mathbf{\hat U}_0^{\text{new}}, (\hat{\sigma}_0^2)^{\text{new}}, \bm{\hat \rho}^{\text{new}}, (\hat{\bm \sigma}^2)^{\text{new}} )$ are obtained by 
\begin{align}
\bm{\hat \Theta}^{\text{new}}	=\mbox{argmax}_{\bm \Theta} \bar \ell(\bm \Theta).
 \label{equ:m_step}
\end{align}

In  Theorem \ref{theorem:closed_form_expression_em}, 
we show that all parameters  
have closed-form expressions in the M-step in (\ref{equ:m_step}), thus avoiding numerically optimizing high-dimensional parameters in the algorithm. {The proof can be found in Supplementary materials SM3.4. Since the M-step is solved exactly via these closed-form expressions, the log-likelihood function is guaranteed to be nondecreasing at each iteration and the algorithm converges to a stationary point \cite{wu1983convergence}. }

\begin{theorem}
\label{theorem:closed_form_expression_em}
Given the parameters estimated from the last iteration, one can compute
$\{s_l(t)\}_{t=1}^n, \{S_l(t)\}_{t=1}^n$ and $\{\tilde{S}_l(t)\}_{t=1}^{n-1}$ by the KF and RTS smoother, for $l=1,\dots, d$.
The
closed-form expressions of $\bm{\hat \Theta}^{\text{new}}$ from the M step, as defined in (\ref{equ:m_step}), are given below.  
\begin{enumerate}
    \item 
    Update $ \hat{\mathbf U}_0^{\text{new}}$ by  
   \begin{align}
    \hat{\mathbf U}_0^{\text{new}} = \tilde{\mathbf V} \tilde{\mathbf U}^T,
    \label{equ:estimate_U0}
  \end{align}
where  $ \tilde{\mathbf U}$ and $\tilde{\mathbf V}$  are the left  and right  singular vectors of the  $\mathbf{\hat{\mathbf Z}} \mathbf Y^T$, respectively. 
   \item 
   Update $(\hat{\sigma}_0^2)^{\text{new}}$ by 
   \begin{align}
       ({\hat \sigma}^2_0 )^{new}
        = \frac{\tr(\mathbf Y^T \mathbf Y)-2\tr(\mathbf Y^T  \mathbf{\hat U}_0^{\text{new}}\hat{\mathbf Z}) + \sum_{l=1}^d \sum_{t=1}^n(s^2_l(t)+S_l(t))}{nk}.
        \label{equ:em_update_sigma_0} 
   \end{align}
    \item   Update $\hat{\rho}_l^{new}$ by solving the following cubic equation in the interval $(-1, 1)$ which has a unique root in  $(-1, 1)$ if $\hat \sigma^2_0>0$ and $\hat \sigma^2_l>0$: 
    \begin{align}
        \label{equ:rho_cubic_equation}
        \beta_0 + \beta_1\rho_l + \beta_2\rho_l^2 + \beta_3\rho_l^3 = 0,
    \end{align}
    where $\beta_0 = n\Bigl(\sum_{t=2}^n {s}_l(t-1) {s}_l(t)+\sum_{t=1}^{n-1}\tilde{S}_l(t)\Bigr)$, $\beta_1=-\sum_{t=1}^n {s^2_l}(t)-\sum_{t=1}^n S_l({t}) - n\sum_{t=2}^{n-1}{s^2_l}(t) - n\sum_{t=2}^{n-1}S_l({t})$, $\beta_2 = (2-n)(\sum_{t=2}^n {s}_l(t-1) {s}_l(t)+\sum_{t=1}^{n-1}\tilde{S}_l(t))$, $\beta_3 = (n-1)(\sum_{t=2}^{n-1}{s^2_l}(t) + \sum_{t=2}^{n-1}S_l(t))$. 
    \item Update $(\hat{\sigma}_l^2)^{\text{new}}$ by
    \begin{align}
    \label{equ:em_update_sigma}
    ({\hat \sigma}_l^2)^{\text{new}} 
     &= \frac{1}{n}\left\{(1-(\hat{\rho}_l^{new})^2){s_l^2(1)} + \sum_{t=2}^n(s_l(t)-\hat{\rho}_l^{new} s_l(t-1))^2 \right\}+    \nonumber\\
     & \qquad \frac{1}{n}\left\{\sum_{t=1}^nS_l(t)+(\hat{\rho}_l^{\text{new}})^2\sum_{t=2}^{n-1}S_l(t)- 2\hat{\rho}_l^{\text{new}}\sum_{t=1}^{n-1}\tilde{S}_l(t)\right\}.
    \end{align}
    \item  Update $\{s_l^{\text{new}}(t)\}_{t=1}^n, \{S_l^{\text{new}}(t)\}_{t=1}^n$ and $\{\tilde{S}^{\text{new}}_l(t)\}_{t=1}^{n-1}$ by KF and RTS smoother using $\mathbf{\hat \Theta}^{\text{new}}$.
    Set $\mathbf{\hat z}_l^{\text{new}}=[s_l^{\text{new}}(1),\dots, s_l^{\text{new}}(n)]^T$ and $\mathbf{\hat Z}^{\text{new}}=[\mathbf{\hat z}_1^{\text{new}},\dots, \mathbf{\hat z}_d^{\text{new}}]^T$.
\end{enumerate}
\end{theorem}

The FMOU in Theorem \ref{theorem:closed_form_expression_em} is nontrivial to derive, and it differs from the conventional EM algorithm for vector autoregressive (VAR) models \cite{shumway1982approach}. First, FMOU avoids the costly matrix inversion required in each step in the KF due to the use of an orthogonal latent factor loading matrix that can be estimated from data. Second, it enables dimension reduction. Third, according to Lemma \ref{lemma:OU_stationary_conditions}, the FMOU algorithm can be set to be stationary or nonstationary,  while the conventional EM for VAR models cannot guarantee stationarity in estimation.

Algorithm \ref{algorithm:EM_FMOU_est_U} summarizes the proposed FMOU approach. 
The output of Algorithm \ref{algorithm:EM_FMOU_est_U} is the MMLE of the parameters $\bm \Theta^{\text{MMLE}}$ in (\ref{equ:mmle}) and posterior mean of the latent factors given $\bm \Theta^{\text{MMLE}}$, denoted as  $\mathbf Z^{\text{post}}=[\mathbf{z}^{\text{post}}(1),...,\mathbf{z}^{\text{post}}(n)]=\mathbb E[\mathbf Z \mid \mathbf Y, \bm \Theta^{\text{MMLE}}]$.  In our application of slip estimation, other quantities, such as the posterior distribution of the mean of the observations and slips at time $t$ follows 
\begin{align}
    \label{equ:posterior_dist_mean}
   \left( \mathbf m(t) \mid \mathbf Y, {\bm \Theta}^{\text{MMLE}} \right) \sim \mathcal{MN}\left({\mathbf U}_{0}{\mathbf z}^{\text{post}}(t), {\mathbf \Sigma}_m^{\text{MMLE}}(t) \right), \\
   \label{equ:posterior_dist_slips}
   \left( \mathbf{z}_s(t) \mid \mathbf Y, {\bm \Theta}^{\text{MMLE}} \right) \sim \mathcal{MN}\left(\mathbf{G}^T {\mathbf U}_0 \mathbf{D}_0^{-2}{\mathbf z}^{\text{post}}(t), {\mathbf \Sigma}^{\text{MMLE}}_s(t) \right),
\end{align}
where ${\mathbf \Sigma}^{\text{MMLE}}_m(t) = {\mathbf U}_0 \mathbf{D}_S(t) {\mathbf U}_0^T $ and $\mathbf \Sigma^{\text{MMLE}}_s(t) = \mathbf G^T {\mathbf U}_0 \mathbf{D}_0^{-2} \mathbf{D}_S(t) \mathbf{D}_0^{-2} {\mathbf U}_0^T \mathbf G$, with $\mathbf{D}_S(t)$ being a $d\times d $ diagonal matrix where the $l$th element is $S_l(t)$ computed by plugging in the MMLE of the parameters for $l=1,2,\dots, d$. Here $\mathbf U_0$ can be either fixed or estimated. In the application of slip estimation,  $\mathbf U_0$ and $\mathbf{D}_0$ are the first $d$ left singular vectors and the largest $d$ singular values of the matrix of Green's function $\mathbf G$, respectively and variance of the noise is usually available from GPS measurements, with details discussed in Section SM3.5. 
Furthermore, in our default setting the initialization of $\mathbf U_0$ and $\sigma^2_0$ are initialized as  
{the first $d$ left singular vectors of the output matrix and $\log(1.5)$, respectively,}
as they do not have a large impact on estimating the mean due to the flexibility of the models with a large number of parameters. { Other initialization ways can be used. For instance, the initialization of $\mathbf U_0$ can be sampled from the Stiefel manifold or specified as a $k\times d$ matrix where the diagonal elements of the first $d$ columns are 1, and 0 otherwise, with $d\leq k$.}

\begin{algorithm}[t]
\caption{Fast EM algorithm for multivariate Ornstein-Uhlenbeck process} 
\begin{algorithmic}[1]

\Input{ $k \times n$ observation matrix $\mathbf{Y}$, time step $1,2,\dots, n$ and selected number of latent processes $d$. The representer $\mathbf{G}$ and noise level $\sigma_0^2$ should be provided if they are fixed.}

\State (Optional) Initialize $\mathbf U_0$ and $\sigma_0^2$ if they are unspecified.
\State Initialize $(\bm{\hat \rho}, \hat{\bm \sigma}^2)$ and then apply KF and RTS smoother to get $\{s_l(t)\}_{t=1}^n, \{S_l(t)\}_{t=1}^n$ and $\{\tilde{S}_l(t)\}_{t=1}^{n-1}$ for $l=1,\dots, d$.
\While {The convergence criteria were not met and the number of iterations is less than the upper bound}

\State (Optional.) If $\mathbf U_0$ is unknown, update $\mathbf {\hat U}_0$ using  Equation (\ref{equ:estimate_U0}). 
\State (Optional.) If $\sigma^2_0$ is unknown, update $\hat \sigma^2_0$ using  Equation (\ref{equ:em_update_sigma_0}).

\State Update $\hat \rho_l$ in the correlation matrix $\mathbf R_l$ for $l=1,\ldots,d$ using  Equation (\ref{equ:rho_cubic_equation}).

\State Update $\hat \sigma_l^2$ for $l=1,\ldots,d$ using  Equation (\ref{equ:em_update_sigma}).

\State Update $\{s_l(t)\}_{t=1}^n, \{S_l(t)\}_{t=1}^n$, $\{\tilde{S}_l(t)\}_{t=1}^{n-1}$ and $\mathbf{\hat Z}$ by KF and RTS smoother.
     
\EndWhile

\Output{$\mathbf{\Theta}^{\text{MMLE}}$ and $\mathbf{Z}^{\text{post}}$.
}
\end{algorithmic}
\label{algorithm:EM_FMOU_est_U}
\end{algorithm}

Several advantages of this EM algorithm are worth mentioning. 
First, the kernel parameters in Gaussian processes are notoriously hard to estimate numerically, and we have $d$ sets of kernel and scale parameters, where $d$ can be on the order of $10^3$ or even $10^6$ in some applications. The estimates of these parameters have closed-form expressions in our EM algorithm, which had not been derived before. Our algorithm substantially improves estimation stability compared to optimization algorithms that numerically optimize in a high-dimensional parameter space. Second, when  $\mathbf U_0$ is unknown and the parameters are distinct in each latent process, originally one needs to solve a constrained optimization problem in the Stiefel manifold \cite{wen2013feasible} to  optimize $\mathbf U_0$ in each of the numerical iteration for estimating the parameters of the model in (\ref{equ:FMOU}) and (\ref{equ:OU}) \cite{gu2020generalized}. In contrast, we derived closed-form expressions for estimating $\mathbf U_0$ in each iteration of the EM algorithm, and hence the algorithm is much faster. Third, we integrate orthogonal projection in the KF and RTS smoother to bypass the matrix inversion operation at each time step.  For $M$ iterations of the EM algorithms, the overall complexity of the parameter estimation process is $\mathcal{O}(Mknd)+\mathcal{O}(Mkd^2)$ when the coefficient $\mathbf U_0$ is unknown, as shown in Table \ref{tab:comparison}.  For the scenario where  $\mathbf U_0$ is derived from the Green's function, we need to perform SVD to obtain $\mathbf U_0$ only once, which has the order $\mathcal O(min({k}^2 k^{\prime}, {k} (k^{\prime})^2))$. 

\subsection{Estimation of the number of latent processes}
\label{subsec:dim_selection}

Estimating a suitable number of factors is crucial to distinguish signals from noise. Various approaches have been developed based on, for instance, information criteria \cite{bai2002determining}, the cumulative percentage of variance explained by the factors  \cite{hotelling1933analysis},  and the ratio of neighboring eigenvalues of the empirical autocovariance matrix \cite{lam2011estimation}. In our scientific application, the goal is not to select the number of factors, but to provide a suitable model for linking the complex slip propagation process to ground deformation.
To enable the variability in the signal to be properly explained by our  model, we select the number of latent factors, $d$, by minimizing the difference between the estimated and measured noise variance:
\begin{equation}  \hat{d} = \argmin_{d} |  ({\sigma}^{\text{MMLE}}_0(d))^2 - \sigma_0^2 |, 
    \label{equ:d_est_known_var}
\end{equation}
where $ \sigma_0^2 $ is the measured variance of the noise from GPS, and $({\sigma}^{\text{MMLE}}_0(d))^2$ is the estimated variance of the noise with $d$ latent factors. We call this approach  variance matching (VM).

When the variance of the noise is unknown, we follow \cite{bai2002determining} to use an information criterion (IC) in Equation (\ref{equ:d_est_PCA_IC}) 
to select the number of factors, which  has the form below: 
\begin{equation}
   \hat d = \argmin_{d} \left\{ \log( ||\mathbf Y - \tilde{\mathbf U}_{1:d}\tilde{\mathbf U}_{1:d}^T \mathbf Y ||_{F}) + C(k, n) \right \},
    \label{equ:d_est_PCA_IC}
\end{equation}
where $|| \cdot ||_F$ denotes the Frobenius matrix norm, $\tilde{\mathbf U}_{1:d}$ is the first $d$ columns from the left unitary matrix of the singular value decomposition of $\mathbf Y $ and $C(k,n)$ is a penalty term, such as  $C(k,n)=d\left(\frac{k+n}{kn}\right)\log(\frac{kn}{k+n})$. This approach will be compared in 
Section \ref{sec:simulations}.

\section{Improved computational scalability and efficiency} 
\label{sec:connections}

We discuss the connection to  other approaches, including the dynamic mode decomposition  \cite{schmid2010dynamic}, the conventional EM algorithm for vector autoregressive models \cite{shumway1982approach}, 
and the generalized probabilistic principal component analysis \cite{gu2020generalized} in Sections  \ref{sec:connection_DMD}-\ref{sec:connection_gppca}, respectively. We  also compare the network inversion filter (NIF) \cite{segall1997time} and its modification \cite{bartlow2011space} for slip estimation in Section 
SM4 in  supplementary materials.  The computational order of different approaches is provided in Table \ref{tab:comparison}. 

\begin{table}[t]
\centering
\begin{tabular}{lcccc}
 \toprule
Methods & computational order & noise models  &  explicit estimators of $\mathbf U_0$ \\ 
 \midrule
FMOU  & $\mathcal{O}(Mknd)+\mathcal{O}(Mkd^2)$  &  Yes & Yes  \\ 
DMD  & $\mbox{min}(\mathcal{O}(kn^2),\mathcal{O}(k^2n))$  &  No & Yes  \\ 
 EM-VAR(1)& $\mathcal{O}(M^* k^3n)$ & Yes& /\\
GPPCA &  $\mathcal{O}(M_1M_2knd)+\mathcal{O}(M_1M_2kd^2)$& Yes& No\\
NIF & $\mathcal{O}(\tilde M k^3n)$ & Yes& /\\

\bottomrule
\end{tabular}
\caption{Comparison of different approaches by the computational order, whether a noise model is included, and whether a closed-form expression in estimating the factor loading matrix is available when the factor processes have distinct correlation parameters. For FMOU, we assume $M$ iterations in EM, where $M\approx 10$s often achieve good performance. For EM-VAR(1), $M^*$ is the number of E-M iterations, and the order is when $d=k$, which is the default setting in \cite{astsa}. For DMD, we only consider the cost of obtaining the first $d$ eigenvectors, whereas obtaining the transition coefficient matrix requires extra $O(k^2d)$ operations.  When each factor contains distinct correlation parameters,  GPPCA requires  $M_1$  iterations and each requires $M_2$  steps to optimize the factor loading matrix in the Stiefel manifold. To estimate the slip, an additional SVD operation of a $k\times k'$ matrix $\mathbf G$ is needed, and it only needs to be done once. 
Thus, the order of FMOU is $\mathcal O(min({k}^2 k^{\prime}, {k} (k^{\prime})^2))+\mathcal O(Mknd)$ and the order of NIF is $\mathcal O(min({k}^2 k^{\prime}, {k} (k^{\prime})^2))+\mathcal O(\tilde Mk^3n)$ with $\tilde M$ being the number of iterations in NIF for slip estimation. 
}
\label{tab:comparison}
\end{table}

\subsection{The FMOU approach extends the dynamic mode decomposition with a symmetric transition matrix for noisy data}
\label{sec:connection_DMD}

The dynamic mode decomposition (DMD) is a popular approach in computational mathematics to obtain the reduced-rank representation of nonlinear dynamical systems \cite{schmid2010dynamic,tu2014dmd}. In DMD, one approximates a real-valued output of $k$ dimensions at time $t$ through $\mathbf{y}(t) \approx \mathbf A \mathbf{y}(t-1)$, where $\mathbf A$ is a $k\times k$ linear transition matrix, estimated by

\begin{align}
\label{equ:dmd_objective_func}
   \mathbf{\hat A} = \argmin_{\mathbf A}\lVert \mathbf{Y}_{2:n}-\mathbf A \mathbf{Y}_{1:n-1} \rVert = \mathbf{Y}_{2:n}\mathbf{Y}_{1:n-1}^+,
\end{align}
where $\mathbf Y_{2:n}=[\mathbf y(2),\dots, \mathbf y(n)]$, $\mathbf Y_{1:n-1}=[\mathbf y(1),\dots, \mathbf y(n-1)]$, $\lVert \cdot \rVert$ stands for the $L_2$ or Frobenius norm, and $\mathbf{Y}_{1:n-1}^+$ represents the Moore-Penrose pseudo-inverse of $\mathbf{Y}_{1:n-1}$.

Estimating the leading eigenpairs of $\mathbf A$ is crucial for low-rank approximation of the system. In \cite{tu2014dmd}, the authors proposed the exact DMD algorithm, which computes the nonzero eigenvalues and the corresponding eigenvectors of $\mathbf A$ with  $\min(\mathcal{O}(k n^2),\mathcal{O}( nk^2))$ operations. The exact DMD algorithm is summarized in Section SM4.1 
in the supplementary materials.

A probabilistic model of the DMD was introduced in \cite{gu2024probabilistic}, which shows the estimation of $\mathbf A$ in DMD is equivalent to the maximum likelihood estimator of $\mathbf A$ in the vector autoregressive model $\mathbf y(t) = \mathbf A \mathbf y(t-1) + \bm{\mathbf{\tilde w}}(t)$,
where $\mathbf{\tilde w}(t) \sim \mathcal{MN}(\mathbf 0, \mathbf{\Sigma}_{\mathbf{\tilde w}})$ represents the innovation with a positive definition matrix $\mathbf{\Sigma}_{\mathbf{\tilde w}}$.   However, both DMD and its probabilistic model assume noise-free observation, which is restrictive in practice.

The FMOU model includes an additional level of noise modeling for DMD with a symmetric transition matrix. To see this, for the FMOU model in 
(\ref{equ:FMOU}) and  (\ref{equ:OU}), denoting $\bar {\mathbf z}(t)=\mathbf U_0 \mathbf z(t)$, the model can be equivalently  written as 
\begin{align}
\mathbf y(t)&=\mathbf {\bar z}(t)+\bm{\epsilon}(t), \label{eq:PDMD_1}\\
\mathbf {\bar z}(t)&=\mathbf A \mathbf {\bar z}(t-1)+\bar {\mathbf w}(t),\label{eq:PDMD_2} 
\end{align}
where $\mathbf A= \mathbf U_0 \bm \Lambda_\rho \mathbf U^T_0$ with $\mathbf U_0 $ being an $k\times d$ orthogonal matrix and $\bm \Lambda_\rho$ being an $d\times d$ diagonal matrix having the $l$th diagonal entry $\rho_l$, and $\bar {\mathbf w}(t) \sim \mathcal N(\mathbf 0, \, \bm \Sigma_{\bar{\mathbf w}})$ with $\bm \Sigma_{\bar{\mathbf w}}= 
\mathbf U_0 \mathbf D_{\sigma} \mathbf U^T_0$ being a $k\times k$ matrix, and $ \mathbf D_{\sigma}=\text{diag}(\sigma_1^2,\dots, \sigma_d^2)$. By multiplying $\mathbf U_0^T$ on both sides of Equations (\ref{eq:PDMD_1})-(\ref{eq:PDMD_2}), we obtain the FMOU model.

Thus, the FMOU approach provides a fast solution of a noise-inclusive, probabilistic DMD model with a symmetric transition matrix. This extension drastically improves the DMD when the data contain noise, as will be numerically demonstrated in Section \ref{sec:simulations}. The computational order of DMD and FMOU is summarized in Table \ref{tab:comparison}, and both are fast in practice. When both $k$ and $n$ are large, the FMOU can be faster than the DMD. 

\subsection{The FMOU approach accelerates the EM algorithm for vector autoregressive models and enables dimension reduction}
\label{subsec:EM_VAR_1}

{The EM algorithm was developed for vector autoregressive models \cite{shumway1982approach}: 
\begin{align}
\mathbf y(t)&= \mathbf G_t \mathbf z(t)+\bm \epsilon(t), \label{eq:y_EM_VAR_1}\\
\mathbf z(t)&= \bm \Phi \mathbf z(t-1) +\mathbf w(t),\label{eq:z_EM_VAR_1}
\end{align}
where $\mathbf G_t$ is a $k\times d$ known factor loading matrix often specified as identity matrix, $\bm \epsilon(t) \sim \mathcal{MN}(\mathbf 0, \bm \Sigma_{\epsilon})$ with covariance $\bm \Sigma_{\epsilon}$, $\bm \Phi$ is an $d\times d$ coefficient matrix and $\mathbf w(t) \sim \mathcal{MN}(\mathbf 0, \bm \Sigma_w)$. Typically, one assumes $\mathbf G_t=\mathbf I_k$ with $k=d$ \cite{astsa}.  The EM algorithm was derived for estimating parameters $\{\bm \Sigma_{\epsilon},  \bm \Phi, \bm \Sigma_w \}$ along with the KF. We call this approach the EM algorithm of the vector autoregressive model of order 1 (EM-VAR(1)). 

The proposed FMOU approach has three advantages. First, we developed fast algorithms to estimate the factor loading matrix and enable selecting a reduced dimension $d$, while an identity factor loading matrix is often assumed with $d=k$ in the EM-VAR(1) approach, when $\mathbf{G}_t$ is unknown. Second,  the KF and RTS smoothers in our FMOU approach only take $\mathcal O(nd)$ operations, which is substantially faster than the EM algorithm as it requires $\mathcal O(nk^3)$ for inverting a $k\times k$ matrix at each time step in KF. Third,  each factor process of the FMOU model can be set to be stationary using Lemma \ref{lemma:OU_stationary_conditions}, yet the EM-VAR(1) approach cannot guarantee that the latent process is stationary. The computational complexity of EM-VAR(1) model \cite{shumway1982approach,astsa}
is given in Table \ref{tab:comparison}.  We will numerically compare FMOU and EM-VAR(1) in Experiment \ref{exp:diffusion_branin}. 
}

\subsection{The FMOU approach enables scalable and robust  estimation for the generalized probabilistic principal component analysis}
\label{sec:connection_gppca}
The FMOU model is connected to the generalized probabilistic principal component analysis (GPPCA) approach \cite{gu2020generalized}, which assumes each latent factor follows a Gaussian process. However, when the variance and range parameters of the latent processes are distinct, in each iteration of the parameter estimation, one needs to apply an iterative algorithm to estimate the factor loading matrix on the Stiefel manifold \cite{wen2013feasible}. In total, one needs $M_1M_2$ iterations, where $M_1$ is the number of iterations in  numerical optimization of the parameters, each requiring $M_2$ iterations to optimize the factor loading matrix. This process can be prohibitively expensive. Furthermore, numerical optimization of $d$ sets of range and variance parameters can also be unstable. In FMOU, we only need $M$ EM iterations, as the estimation of parameters  
have closed-form expressions. 
We use Experiment \ref{exp:supplement_compare_U_and_rmse} to numerically compare the stability and computational cost of FMOU and GPPCA. 

\begin{experiment}
\label{exp:supplement_compare_U_and_rmse}
The data are generated by Equations  (\ref{equ:FMOU})-(\ref{equ:OU}). The orthogonal matrix $\mathbf U_0 \in \mathbb R^{k\times d}$ is sampled from the Stiefel manifold with two configurations with $(k=20, d=5)$ and $(k=40, d=10)$, where $d$ is assumed to be known. The latent factors are generated with inputs $t=1,\ldots,n$ for three scenarios with $n \in \{100, 200, 400\}$. The parameters $\rho_l$ and $\sigma_l^2$ are sampled from $\mbox{Unif } (0.95, 1)$ and $\mbox{Unif }(0.5, 1)$ for $l=1,...,d$, respectively, and the variance of the noise  is  $\sigma_0^2 =0.2$.  We repeat the simulation $N=20$ times under each configuration.
\end{experiment}

\begin{figure}[t]
\centering
\begin{tabular}{c}
\includegraphics[scale=0.55]{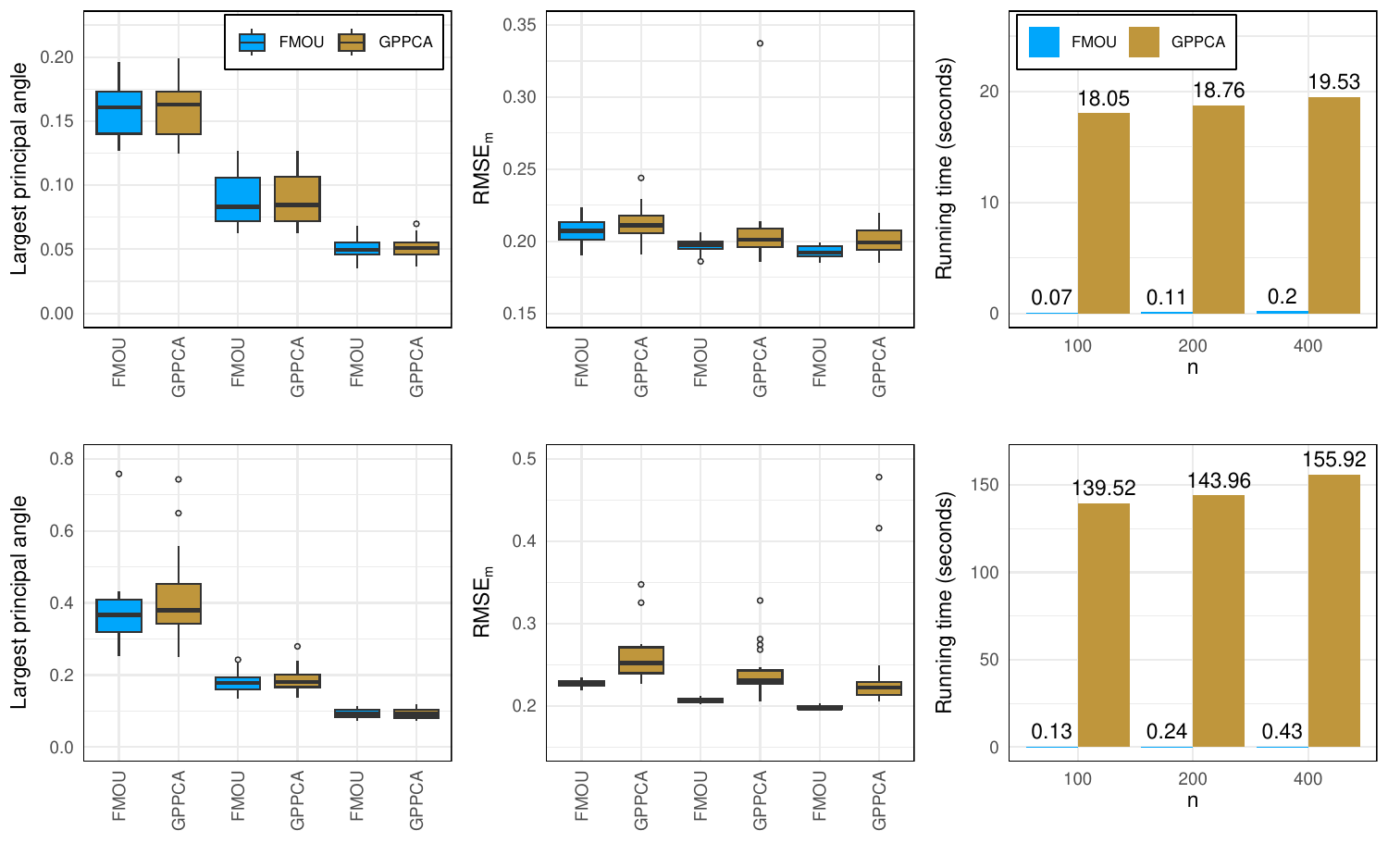}
\end{tabular}
\vspace{-.1in}
\caption{
The largest principal angle between the true loading matrix $\mathbf U_0$ and its estimates, the $\text{RMSE}_m$ and the running time of performing models in Experiment \ref{exp:supplement_compare_U_and_rmse} with correctly specified number of latent factors in FMOU and GPPCA. The first and second rows show the scenarios with $(k=20, d=5)$ and $(k=40, d=10)$, respectively.
The first 2, middle 2, and last 2 boxes are associated with $n=100$, $n=200$, and $n=400$, respectively, in each figure. 
}
\label{fig:compare_FMOU_GPPCA}
\end{figure}

We use two criteria to evaluate the estimation of the latent factor loadings and the mean. First, we compute the largest principal angle $\phi_d$ between $\mathcal{M}(\hat{\mathbf U}_0)$ and $\mathcal{M}({\mathbf U}_0)$,  the linear subspaces spanned by the estimation $\hat{\mathbf U}_0$ and by the truth $\mathbf U_0$, respectively, to measure the closeness of two linear subspaces.  Let $0 \leq \phi_1 \leq \ldots \leq \phi_d \leq \pi/2$ be the principal angles:
\begin{align}
    \phi_l &= \text{arccos}\left(\underset{\mathbf u \in \mathcal{M}(\mathbf U_0), \hat{\mathbf u} \in \mathcal{M}(\hat{\mathbf U}_0)}{\max} |\mathbf{u}^T \hat{\bf u}| \right) = \text{arccos}(|\mathbf{u}_l^T \hat{\mathbf u}_l|), 
    \label{equ:largest_principal_angle}
\end{align}
such that $ \lVert \mathbf u \rVert = \lVert \hat{\mathbf u} \rVert = 1, \mathbf{u}^T \mathbf{u}_l=\hat{\mathbf u}^T \hat{\mathbf u}_l = 0$, with ${\bf u}_l$ and $\hat{\bf u}_l$ being the $l$th column of $\mathbf U_0$ and $\hat{\mathbf U}_0$ for $l=1,\ldots,d-1$. {A smaller largest principal angle indicates a better estimation.}

Second, we compute the root of the mean squared error for the mean of the  observations: 
\begin{align*}
    \text{RMSE}_m &= \frac{1}{N}\sqrt{\frac{\sum_{i=1}^{k}\sum_{t=1}^n \big(\hat{y}_{i}(t)- \mathbb{E}[y_i(t)] \big)^2}{k n}}, \label{equ:rmse_mean}
\end{align*}
where $\hat{y}_i(t)$ is the predictive mean of $y_i(t)$ and $N$ is the number of replications. 

Figure \ref{fig:compare_FMOU_GPPCA} compares FMOU with GPPCA. The largest principal angles between the exact loading matrix and its estimations from FMOU and GPPCA are close. The FMOU is consistently better than the GPPCA in estimating the mean, and the effect is more pronounced for $d=10$, a larger dimension of the latent space. This is because numerically optimizing a large number of parameters and factor loading matrix leads to a larger error from GPPCA, whereas such a problem is overcome by closed-form expressions in each EM iteration in FMOU. Furthermore, the FMOU is much faster than the GPPCA, as shown in the right panels in Figure \ref{fig:compare_FMOU_GPPCA}, as the estimation of factor loadings has a closed-form solution in FMOU, yet numerical optimization in Stiefel manifold is required in each optimization step in GPPCA. 

Hence, FMOU achieves tremendous improvements in terms of scalability and efficiency for the model in  (\ref{equ:FMOU})-(\ref{equ:OU}), compared to  GPPCA. In Sections \ref{sec:simulations} and \ref{sec:real_application}, we use extensive simulated and real examples with correctly specified or misspecified scenarios to compare FMOU with alternative approaches. The  numerical results  are computed by a  macOS Mojave 
system with an 8-core Intel i9 processor running at 3.60 GHz and 32 GB of RAM.  

\section{Simulated experiments}
\label{sec:simulations}
In this section, we numerically compare our method with alternative approaches by simulation. The experiments are split into two subsections. The scenarios with an estimated factor loading matrix are considered in Section \ref{subsec:est_factor_loadings}, whereas Section  \ref{subsec:fixed_factor_loadings} focuses on cases with a fixed factor loading matrix. Both correctly specified models and misspecified models are considered. 
In Section \ref{subsec:est_factor_loadings}, we compared with DMD and two data-driven latent factor models, denoted as LY1 and LY5, introduced in \cite{lam2012factor}. In the LY1 and LY5 methods, the number of latent factors $\hat{d}$ is first estimated by minimizing the ratio of neighboring eigenvalues of $\mathbf C:=\sum_{p=1}^{p_0} \bm{\hat \Sigma}_y(p)\bm{\hat \Sigma}^T_y(p)$, with $p_0=1$ for LY1 and $p_0=5$ for LY5, where $\bm{\hat \Sigma}_y(p)$ denotes the sample covariance matrix with a lag time $p$. In both methods, the factor loading matrix $\mathbf U_0$ is estimated by the first ${d}$ eigenvectors of $\mathbf C$ corresponding to the largest ${d}$ eigenvalues. Results with an estimated latent factor loading matrix $\hat d$ are shown in Section SM5 
in supplementary materials. {For Experiment 3 in Section \ref{subsec:est_factor_loadings}, we also include EM-VAR(1) \cite{shumway1982approach} with the default setting $\mathbf G_t=\mathbf I_k$ in Equation (\ref{eq:y_EM_VAR_1}) for estimating the mean from noisy observations. The non-orthogonal eigenvectors of the $\Phi$ in Equation (\ref{eq:z_EM_VAR_1}) play a similar role as the factor loading matrix in the FMOU model, which is estimated by the data. We do not include EM-VAR(1) in other simulated examples as they are too expensive to compute. } In  Section \ref{subsec:fixed_factor_loadings}, we construct two additional experiments where the factor loadings are either sampled from the  Stiefel manifold or derived from the singular vectors of the Green's function used in real data, mimicking the slip propagation process.  
We compare FMOU with NIF \cite{segall1997time} for these two experiments. 
Other than estimation error and computational time, we also record the proportion of the signals and slips covered in $95\%$ posterior credible intervals and the average length of the intervals of the FMOU approach in Tables SM2-SM6 in supplementary materials, yet these uncertainty measures are not available for other methods, such as DMD. 

\subsection{Simulated experiments with estimated factor loadings}
\label{subsec:est_factor_loadings}
We first study Experiment \ref{exp:compare_U_and_rmse}, where all parameters, $(\mathbf U_0,\bm  \sigma^2_0, \bm \sigma^2, \bm \rho )$, are estimated. 

\begin{experiment}[Correctly specified models with estimated factor loading matrices]

\label{exp:compare_U_and_rmse}
  
The data are generated by Equations  (\ref{equ:FMOU})-(\ref{equ:OU}) with the orthogonal matrix $\mathbf U_0 \in \mathbb R^{k\times d}$  sampled from the Stiefel manifold with $k=20$ and $d=5$. Here $\rho_l$ and $\sigma_l^2$ are sampled from $\mbox{Unif } (0.95, 1)$ and $\mbox{Unif }(0.5, 1)$ for $l=1,...,d$, respectively. Six scenarios with three different numbers of time points $n \in \{100, 200, 400\}$ and two unobserved variances of the noise $\sigma_0^2 \in \{1,2\}$ are considered. We repeat the simulation $N=20$ times for each scenario. 

\end{experiment}

\begin{figure}[t]
\centering
\begin{tabular}{c}
\includegraphics[scale=0.55]{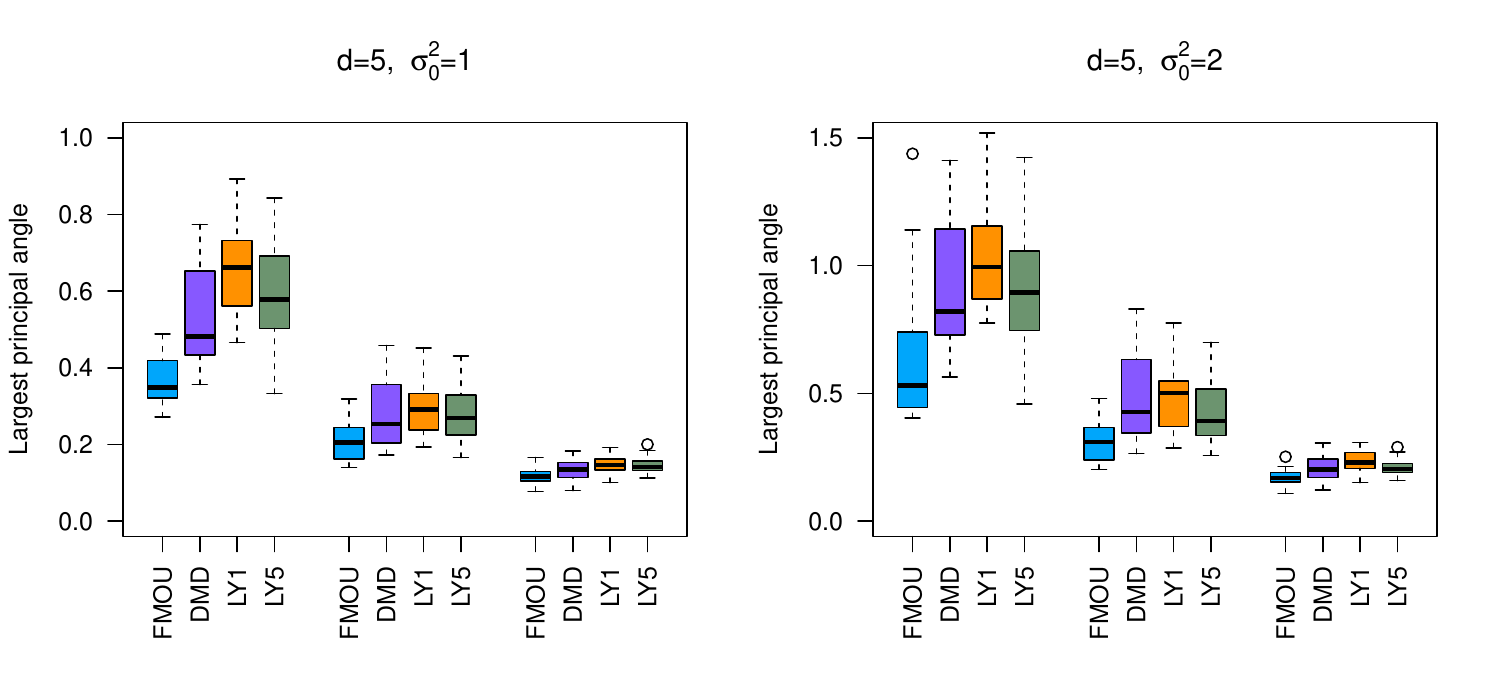}
\end{tabular}
\vspace{-.1in}
\caption{
The largest principal angle (from 0 to $\pi/2$) between the true loading matrix $\mathbf U_0$ and its estimates from 4 methods in Experiment \ref{exp:compare_U_and_rmse} with correctly specified latent factors. 
The variance of the noise is assumed to be $\sigma^2_0=1$ and $\sigma^2_0=2$ for the left and right panels, respectively. 
In each panel, the first 4, middle 4, and last 4 boxes are associated with a different number of time points, $n=100$, $n=200$, and $n=400$, respectively.}
\label{fig:compare_U}
\end{figure}

In Figure \ref{fig:compare_U}, we show the largest principal angle of factor loading matrix $\mathbf U_0$ for all approaches with a correctly specified $d$. Across all scenarios, the estimation by the FMOU approach has the smallest principal angles between the estimated and true  factor loading matrix.  In Figure 
{SM1} in the supplementary materials, we show that the number of latent factors can be correctly estimated by the IC in Equation (\ref{equ:d_est_PCA_IC}), and the estimation is more accurate than the alternatives.

\begin{table}[t]
\centering
\begin{tabular}{lccccccc}
 \toprule
$d=5$ &  & $\sigma_0^2=1$ & & & & $\sigma_0^2=2$ & \\ 
 & $n=100$ & $n=200$ & $n=400$ & &$n=100$ & $n=200$ & $n=400$ \\ 
 \midrule
FMOU  & \textbf{0.38} & \textbf{0.35} & \textbf{0.33} & & \textbf{0.50} & \textbf{0.44} & \textbf{0.41} \\  
DMD  & 0.63 & 0.63 & 0.63 & &0.77& 0.76 & 0.77\\ 
LY1  & 0.91 & 0.58 & 0.57 & & 1.3 & 0.83 & 0.81 \\ 
LY5  & 0.89 & 0.57 & 0.58 & & 1.3& 0.81& 0.81 \\
\bottomrule
\end{tabular}
\caption{Average of $\text{RMSE}_m$ over $N$ repeats of Experiment \ref{exp:compare_U_and_rmse} with the truth $d=5$. {The average standard deviations of the output across all sample sizes and repeats are 2.90 and 3.10 for $\sigma_0^2=1$ and $\sigma_0^2=2$, respectively. } 
}
\label{table:unknown_U_rmse}
\end{table}

\begin{figure}[t]
    \centering
    \includegraphics[scale=0.5]{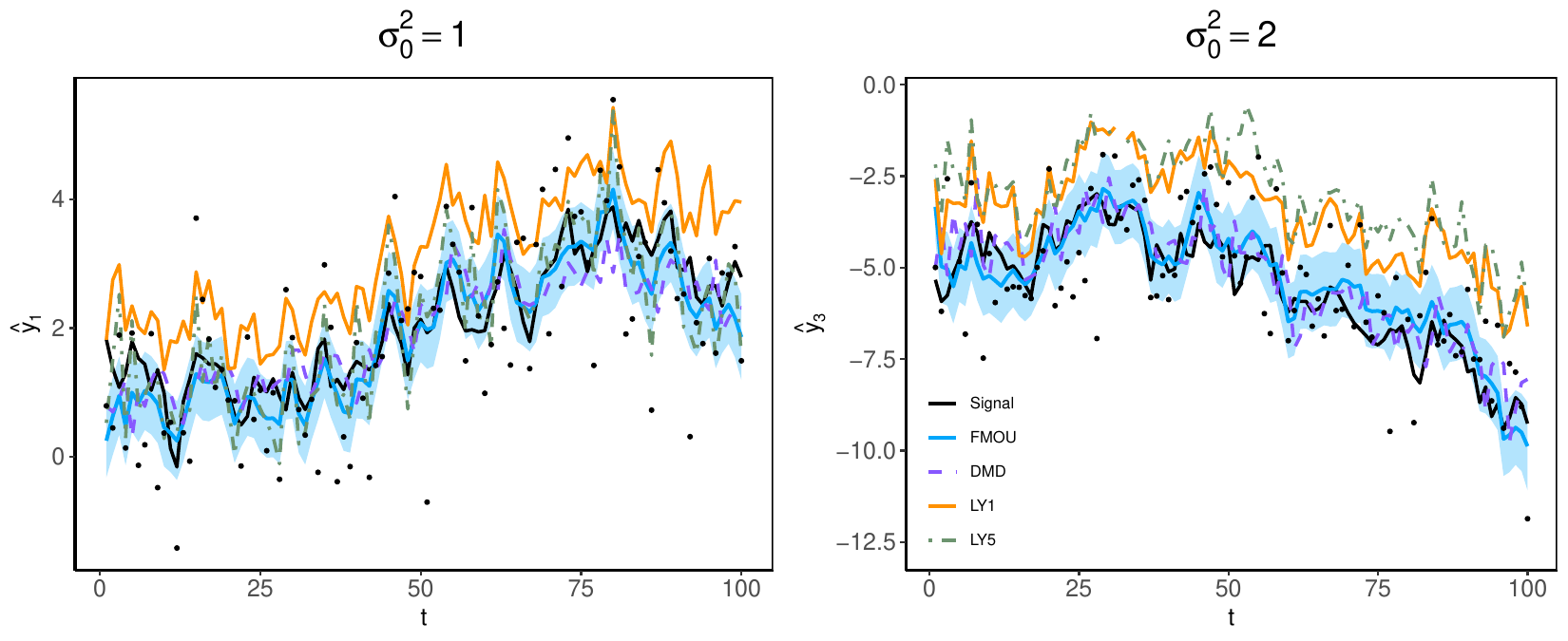}
    \vspace{-.1in}
    \caption{Predictive mean by FMOU (solid blue curves), DMD (dashed purple curves), LY1 (solid orange curves) and LY5 (dashed green curves), from one repetition in Experiment \ref{exp:compare_U_and_rmse}. The observations and mean of the observations are plotted by the black circles and curves, respectively. The blue-shaded area is the 95\% posterior credible interval given by FMOU. 
    }
    \label{fig:unknown_U_pred_mean}
\end{figure}

The RMSE$_m$ in estimating the mean of the data in Experiment \ref{exp:compare_U_and_rmse} is shown in Table \ref{table:unknown_U_rmse}. 
The FMOU achieves higher accuracy under all combinations with different $n$ and $\sigma_0^2$. In Figure \ref{fig:unknown_U_pred_mean}, we plot the observations, the mean, and estimation by different methods for $\sigma^2_0=1$ and $\sigma^2_0=2$. 
The predictive mean from FMOU is close to the truth, plotted as black curves, and the $ 95\%$ posterior credible interval of the mean covers the truth most of the time. {Based on Equation  (\ref{equ:posterior_dist_mean}), the $95\%$ posterior credible interval of the $i$th observation mean at time $t$ is given by $[(\mathbf U_0\mathbf z^{\text{post}}(t))_i -1.96\Sigma^{\text{MMLE}}_{m,ii}(t), (\mathbf U_0\mathbf z^{\text{post}}(t))_i + 1.96\Sigma^{\text{MMLE}}_{m,ii}(t)]$, where $(\mathbf U_0\mathbf z^{\text{post}}(t))_i$ is the $i$-th element of $\mathbf U_0\mathbf z^{\text{post}}(t)$ and  $\Sigma^{\text{MMLE}}_{m,ii}(t)$ is the $i$-th diagonal element of $\mathbf{\Sigma}^{\text{MMLE}}_{m,ii}(t)$.}

\begin{experiment}[Misspecified models with estimated factor loading matrices] 
\label{exp:diffusion_branin}
We consider two  test cases where signals are generated by physical processes different from our  models.
\begin{enumerate}[label=(3\alph*)]
    \item \textbf{Linear diffusion} \cite{carslaw1906introduction}. 
The signal is governed by the partial differential equation $\frac{\partial u(x,t)}{\partial t} = D\frac{\partial^2 u(x,t)}{\partial x^2}$, where $u(x,t)$ represents the concentration of the diffusing material at location $x$ and time $t$, and $D$ is the diffusion coefficient. We follow \cite{lu2020prediction} to let $D=1$, discretize the spatial domain $[0,1]$ into {$300$} equally spaced grid points, $u(x,0)=0$, and let a boundary condition be applied at one end with a constant external concentration of 1. The signal is generated over a time interval $t\in[0,0.2]$ using {$n=300$} with a numerical solver \cite{soetaert2010solving}. Three noise variances {$\sigma_0^2 \in \{0.01^2,0.05^2, 0.3^2\}$}  
are tested.
    \item \textbf{Branin function} \cite{picheny2013benchmark}. The underlying signal is generated by the Branin function, which takes a two-dimensional input $(x_1,x_2)$ and yields a scalar output given by $f(x_1,x_2)=a(x_2-bx_1^2+cx_1-r)^2+s(1-t)\cos(x_1)+s$, where $x_1\in [-5,10]$ and $x_2 \in [0,15]$. The recommended parameter values are used: $a=1, b=\frac{5.1}{4\pi^2},c=\frac{5}{\pi}, r=6, s=10$ and $t=\frac{1}{8\pi}$. When generating the signal, the input domain is uniformly discretized into a $300\times 300$ grid. Noisy observations are then obtained via $y(x_1, x_2)=f(x_1,x_2)+\epsilon$, where $\epsilon$ represents independent Gaussian noise with variance $\sigma^2_0$. We test the effects of varying noise variances, specifically $\sigma_0^2 \in \{1, 5^2,20^2\}$.
\end{enumerate}
\end{experiment}

\begin{figure}[t]
    \centering
    \includegraphics[scale=0.65]{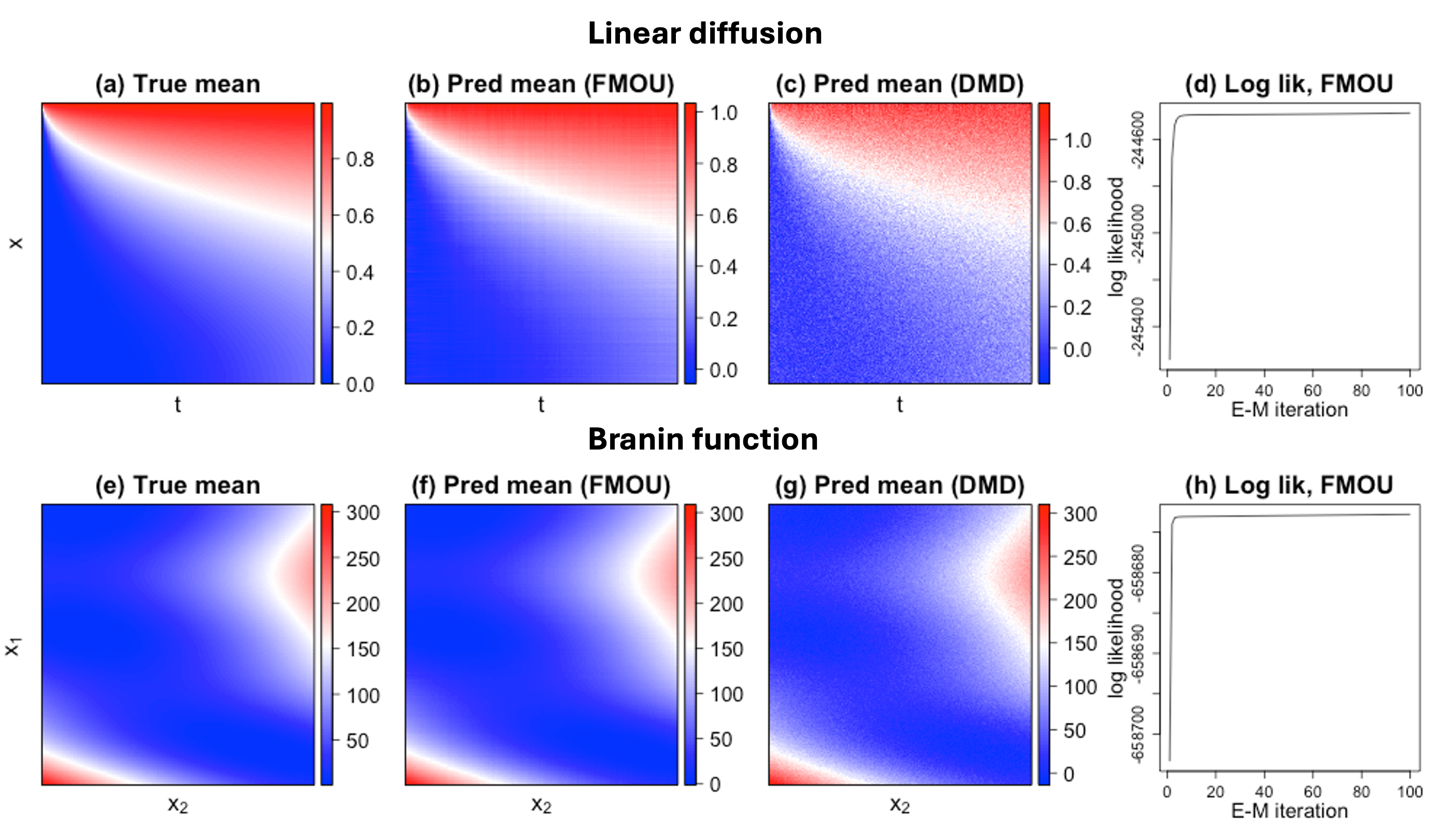}
    \vspace{-.1in}
    \caption{{(a) True mean generated by linear diffusion. (b) Predictive mean by FMOU, where observations are generated with $\sigma_0^2=0.05^2$.  (c) Predictive mean by DMD (i.e., $\hat{u}_\text{DMD}(x,t)$) where observations are generated with $\sigma_0^2=0.05^2$. (d) The convergence of the log likelihood in the EM algorithm of the FMOU approach. (e) True mean generated by the Branin function. (f) Predictive mean by FMOU, where observations are generated with $\sigma_0^2=25$.(g) Predictive mean by DMD, where observations are generated with $\sigma_0^2=25$. (h) The convergence of the log likelihood in the EM algorithm of the FMOU approach. }}
    \label{fig:diffusion}
\end{figure}

Table \ref{tab:exp2} presents the average $\text{RMSE}_m$ {and computational time} for signal estimation over $N=20$ simulations across various approaches. The results show that FMOU consistently achieves better accuracy under all noise levels, particularly excelling in large-noise scenarios. {Moreover, FMOU is much faster over the EM-VAR(1) method with the default setting \cite{astsa} as the FMOU enables dimension reduction of the latent factors and avoids inverting the covariance matrix of the observations at each time in the Kalman filter. } Figure  \ref{fig:diffusion} provides the comparison between FMOU and {DMD} in estimating the signal from noisy observations generated by the linear diffusion equation and {Branin function}, respectively. The signal estimated by FMOU closely aligns with the ground truth, while DMD exhibits larger deviations. {We highlight that the convergence of the log likelihood in the FMOU approach only takes around 5-10 iterations in EM algorithm shown in Figure \ref{fig:diffusion} (d) and (h).} 

\begin{table}[t]
    \centering
    \setlength{\tabcolsep}{3pt}
    {
    \begin{tabular}{ccccccccc}
     \toprule
    & \multicolumn{4}{c}{Linear diffusion} & \multicolumn{4}{c}{Branin function}  \\
        & RMSE$_m$ & RMSE$_m$  & RMSE$_m$ & Avg & RMSE$_m$ & RMSE$_m$& RMSE$_m$ & Avg \\
    & $\sigma^2_0=0.01^2$ & $\sigma^2_0=0.05^2$ & $\sigma^2_0=0.3^2$ & Time& $\sigma^2_0=1$ & $\sigma^2_0=5^2$ & $\sigma^2_0=20^2$ & Time \\
    \midrule
    FMOU & $\bf{0.0024}$ & $\bf{0.0083}$ & $\bf{0.038}$ & $0.56$ &$\bf{0.14}$ & $\bf{0.60}$ & $\bf{2.2}$ & $\bf{0.23}$ \\
    DMD & $0.010$ & $0.047$ & $0.29$& $\bf{0.44}$ & ${0.79}$ & ${4.5}$ & $19$ & $0.41$\\
    LY1 &$0.10$ & $0.050$ & $0.30$ & $3.6$ & $2.4$ & $11$ & $20$ & $3.5$\\
    LY5 & $0.010$ & $0.050$ & $0.30$ & $18$ &$5.9$ & $5.8$ & $20$ & $17$\\
    {EM-VAR(1)} & {$0.0098$} & {$0.050$} & {$0.30$} & $672$ & {$4.1$} & {$5.5$} & {$60$} & $807$\\
    \bottomrule
    \end{tabular}
    \setlength{\tabcolsep}{6pt}
    }
    \caption{
    {Average $\text{RMSE}_m$ and computational time (in seconds) for  Experiment \ref{exp:diffusion_branin}. $\text{RMSE}_m$ is averaged over $N=20$ repeats per noise level. The computational time is averaged over three noise levels and $N=20$ repeats.} {For linear diffusion, the average standard deviations of the output are 0.29 ($\sigma_0^2=0.01^2$), 0.30 ($\sigma_0^2=0.05^2$) and 0.42 ($\sigma_0^2=0.3^2$). For Branin function, the average standard deviations of output are 51.57 ($\sigma_0^2=1$), 51.81 ($\sigma_0^2=5^2$) and 55.32 ($\sigma_0^2=20^2$). 
    }}
    \label{tab:exp2}
\end{table}

\subsection{Simulated experiments with known factor loadings}
\label{subsec:fixed_factor_loadings}
 
In this section, we assume that $\mathbf U_0$ is given by the SVD  of Green's function. The observations are related to the unobserved slip by Equation (\ref{equ:reconstructed_slip}). 
We quantify the estimated error of slips by $\text{RMSE}_s$:
\begin{align}
    \text{RMSE}_s &= \frac{1}{N}\sqrt{\frac{\sum_{t=1}^n \big(\mathbf{\hat z}_s(t) - \mathbf z_s(t)\big)^T \big(\mathbf{\hat z}_s(t) - \mathbf z_s(t)\big)}{k^{\prime}n}},
\end{align}
where $\mathbf{\hat z}_s(t)=(\hat{z}_{s,1}(t),\dots, \hat{z}_{s,k'}(t))^T$ is the estimated slips at time $t$, computed by $\mathbf{\hat z}_s(t) = \mathbf G^T \mathbf U_0 \mathbf D_0^{-2}\mathbf{\hat z}(t)$ with $\hat{\mathbf z}(t)$ being the estimation of latent factors.

\begin{experiment}[Correctly specified models with a known factor loading matrix]
\label{exp:GP_factor_simulation}

{{The data are generated by Equations (\ref{equ:FMOU})-(\ref{equ:OU}), and the slips are computed by Equation (\ref{equ:reconstructed_slip}), which approximate the data generating system in Equation (\ref{equ:displacement}).}} 
The orthogonal matrices $\mathbf U_0 \in \mathbb R^{k\times k}$, $\mathbf V_0 \in \mathbb R^{k^{\prime} \times {k}}$ are generated from the Stiefel manifold with two configurations: (1) $k=25, k'=150, d=6$ and (2) $k=32, k'=100, d=8$. The $l$th latent factor $\mathbf{z}_l$ is generated with inputs $t=1,\ldots,n$ for $n \in \{100, 200, 300\}$. We have $\mathbf{D}_0 = diag(d_1, \ldots, d_k)$ with $d_l$ sampled from $Unif(0, 1)$ for $l=1,...,{k}$ and rearranged decreasingly. Parameters $\rho_l$ and $\sigma_l^2$ are sampled from $Unif(0.95, 1)$ and $Unif(1, 2)$, respectively, for $l=1,...,d$. The matrix of the Green's function is constructed by $\mathbf G =\mathbf U_0 \mathbf{D}_0 \mathbf V^T_0$. 
The variance of the noise is assumed to be $\sigma_0^2 =1.5$. We repeated simulation $N=20$ times for each configuration. 
\end{experiment}

\begin{figure}[t]
\centering
\includegraphics[scale=0.55]{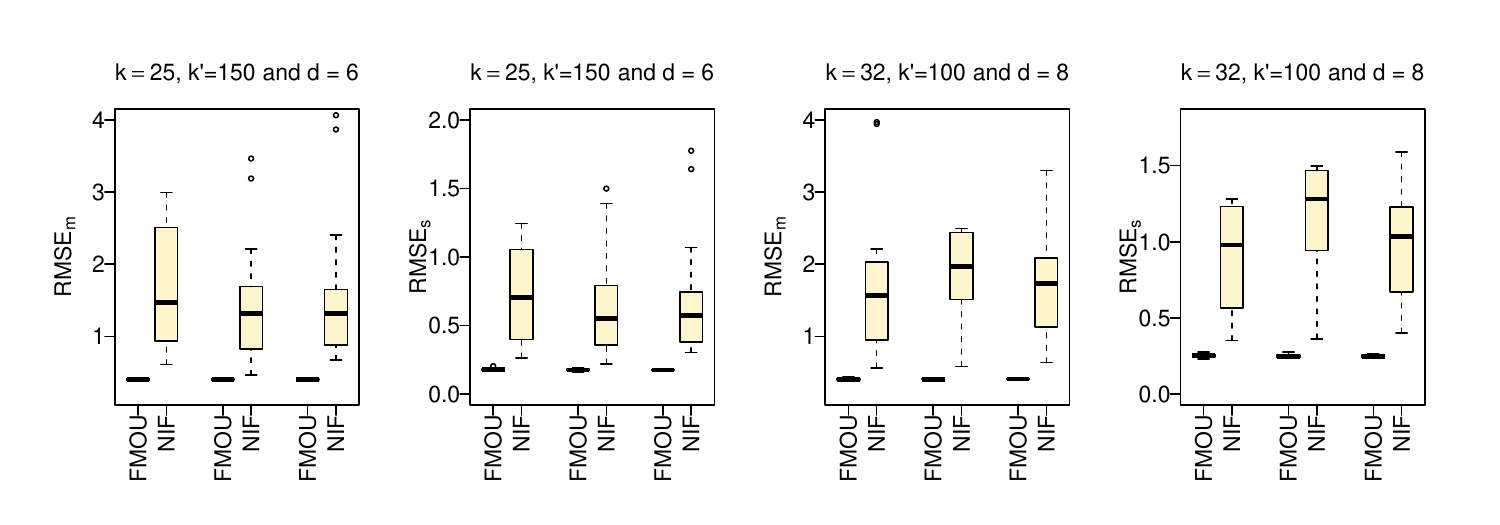}
\vspace{-.2in}
\caption{Box plots of $\text{RMSE}_m$ and $\text{RMSE}_s$ by FMOU and NIF  in Experiment \ref{exp:GP_factor_simulation}. The number of factors $d$ is correctly specified for both FMOU and NIF. In each subfigure, the first 2, middle 2, and the last 2 boxes are based on $n=100$, $n=200$ and $n=300$, respectively. 
}
\label{fig:GP_factors_rmse}
\end{figure}

We compare FMOU with NIF, as other previously compared methods, such as DMD, LY1 and LY5, estimate the factor loading matrix. The $\text{RMSE}_m$ and $\text{RMSE}_s$ by FMOU and NIF are provided in Figure \ref{fig:GP_factors_rmse}. As NIF does not involve the selection of the number of factors $d$, to eliminate the effect of selecting $d$, we assume $d$ is known for both FMOU and NIF. Across all configurations, the FMOU model consistently outperforms NIF in estimating the mean of the observations and the slip. Additionally, Figure SM2  
in supplementary materials compares the estimation of $d$ in Experiment \ref{exp:GP_factor_simulation} and it demonstrates that {IC}    in Equation (\ref{equ:d_est_PCA_IC}) can correctly identify the number of factors with a moderately large number of time points. The improvement by FMOU for Experiment \ref{exp:GP_factor_simulation} is not surprising as data are generated by FMOU model. Next we use a misspecified model to illustrate the flexibility of FMOU estimation.

\begin{experiment}[A misspecified latent factor model for estimating a 2D elliptical slip]
\label{exp:ellipse_simulation}
This simulation uses the Green's tensor $\mathbf G$ inherited from the real data for inferring the slip propagation in the Cascadia zone, where $k=200$, $k^{\prime}=1978$, $n=88$. 
An elliptical slip region is generated on a triangular mesh with a fixed center at $(\xi_{0,1}, \xi_{0,2})=(123.2^{\circ}\text{W}, 46.0^{\circ}$N) and a fixed semi‐minor axis extending from ($123.9^{\circ}$W, $46.0^{\circ}$N) to ($122.5^{\circ}$W, $46.0^{\circ}$N) with ground distance $r_0 = 108$km. The semi-major axis grows along the longitude as $r(t) = r_0/3 + vt$ with the growth rate $v=8$km/day. Data is generated through model (\ref{equ:displacement}) and the slip at the $h$th fault patch $(\xi_{h,1}, \xi_{h,2})$ and time $t$ is given by
\begin{align}
 z_{s,h}(t) = z_{s,max}(1 -  E(\bm \xi_h, t)^2 )\times \mathbbm{1}_{\{ E(\bm \xi_h, t) \leq 1 \}}, 
 \label{equ:ellipse_slip}
 \end{align}
 where $ E(\bm \xi_h, t) = (\xi_{h, 1}-\xi_{0,1})^2/r_0^2+ (\xi_{h, 2}-\xi_{0,2})^2/r(t)^2$, $h=1,\ldots,k^{\prime}, t=1,\ldots, n$.  
The slips have a peak value of $z_{s,max}=3$cm at the center and drop to zero outside the ellipse. We consider three configurations with known noise variances $\sigma_0^2 \in \{0.01^2, 0.05^2, 0.2^2\} cm^2$. 
\end{experiment}

In Experiment \ref{exp:ellipse_simulation}, both $\mathbf U_0$ and $\sigma_0^2$ are known to mimic the application of the real data.  We use the VM method in  (\ref{equ:d_est_known_var}) to estimate the number of latent factors. 
Table \ref{table:ellipse_slips_results} records the numerical comparisons between FMOU and NIF for Experiment \ref{exp:ellipse_simulation}. We consider two variants of NIF: one with $d=k$ and the other one with an estimated $\hat{d}$ from {VM}. 
The FMOU achieves better accuracy in estimating both the mean of the observations and slips. Additionally, FMOU is considerably faster than the NIF model, as computing the likelihood function in FMOU only requires $\mathcal O(knd)$ operations. 
These orders do not consider SVD of Green's function which only needs to be done once. Furthermore, as each step of EM algorithm has closed-form expressions, FMOU is also robust in estimating a large number of parameters.

\begin{table}[t]
\centering
\begin{tabular}{lcccc}
 \toprule
 $\sigma_0^2=0.01^2$ & $\hat{d}$ & Avg. $\text{RMSE}_m$ & Avg. $\text{RMSE}_s$ & Running time (seconds) \\ 
 \midrule
FMOU  & 105 & \boldmath{ $0.0039$} & \boldmath{$0.23$} & {\boldmath{$10$}} \\ 
NIF  & 105 & {$0.11 $}& {$0.64$} & {$549$}\\ 
NIF  & 200 & {$0.11$} & {$0.64$} & {$2271$}\\ 
\midrule
$\sigma_0^2=0.05^2$ & $\hat{d}$ & Avg. $\text{RMSE}_m$ & Avg. $\text{RMSE}_s$ & Running time (seconds) \\ 
 \midrule
FMOU  & 67 & \boldmath{$0.013$} & \boldmath{$0.34$} & {\boldmath{$6.0$}} \\ 
NIF  & 67 & {$0.12$} & {$0.66$} & {$251$}\\ 
NIF  & 200 & {$0.12$} & {$0.66$} & {$2247$}\\ 
\midrule
 $\sigma_0^2=0.2^2$ & $\hat{d}$ & Avg. $\text{RMSE}_m$ & Avg. $\text{RMSE}_s$ & Running time (seconds) \\ 
 \midrule
FMOU  & 30 & \boldmath{$0.034$} & \boldmath{$0.48$} & {\boldmath{$3.6 $}} \\ 
NIF  & 30 & {$0.12$} & {$0.67$} &{$100$}\\ 
NIF  & 200 & {$0.12$} & {$0.66$} & {$2243$}\\ 
\bottomrule
\end{tabular}
\caption{Results for Experiment \ref{exp:ellipse_simulation}. 
The standard deviation of the mean of the output and the slip are 0.19 and 0.85, respectively. 
FMOU takes {9.2, 5.3 and 3.2} seconds for $\sigma_0^2 = 0.01^2, 0.05^2, 0.2^2$ to estimate the number of latent factors, $d$, respectively, and the part of the computation accounts for the majority of the running time from FMOU listed in the table. 
}
\label{table:ellipse_slips_results}
\end{table}

\begin{figure}[t]
\centering
\begin{tabular}{c}
\hspace{-0.125in}
\includegraphics[scale=.43]
{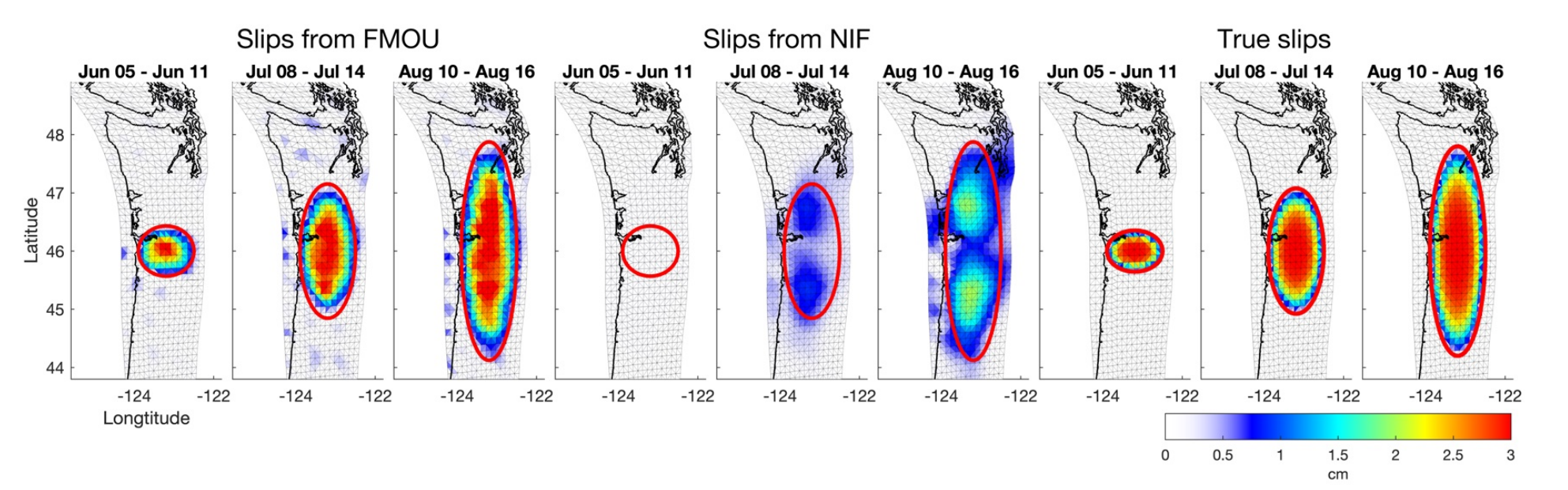}
\end{tabular}
\vspace{-0.22in}
\caption{Seven-day averages of slips estimated by FMOU (left 3 panels) and NIF
(middle 3 panels) in Experiment \ref{exp:ellipse_simulation} when $\sigma_0^2=0.01^2$ $cm^2$. The true slips (right 3 panels) propagate as a growing ellipse. The red boundary shows the front of the slip region. 
}
\label{fig:ellipse_simulation_slip_plot}
\end{figure}

Figure \ref{fig:ellipse_simulation_slip_plot} graphs the 7-day averaged slips in centimeters, such that $\bar{z}_{s,h}(t) = \sum^{6}_{t'=0}\hat{z}_{s,h}(t+t') /7$ for $h=1,\dots, k'$. Both FMOU and NIF detect the front of ellipses and estimate slips' propagation within the elliptical regions. The slips estimated by FMOU model align more closely with the truth which demonstrates that FMOU is more accurate in estimation.

\section{Estimating slip propagation in Cascadia}
\label{sec:real_application}
\subsection{Data  and methods}

\begin{figure}[t]
\centering
\begin{tabular}{cc}
\hspace{-.18in}
\includegraphics[scale=0.38]{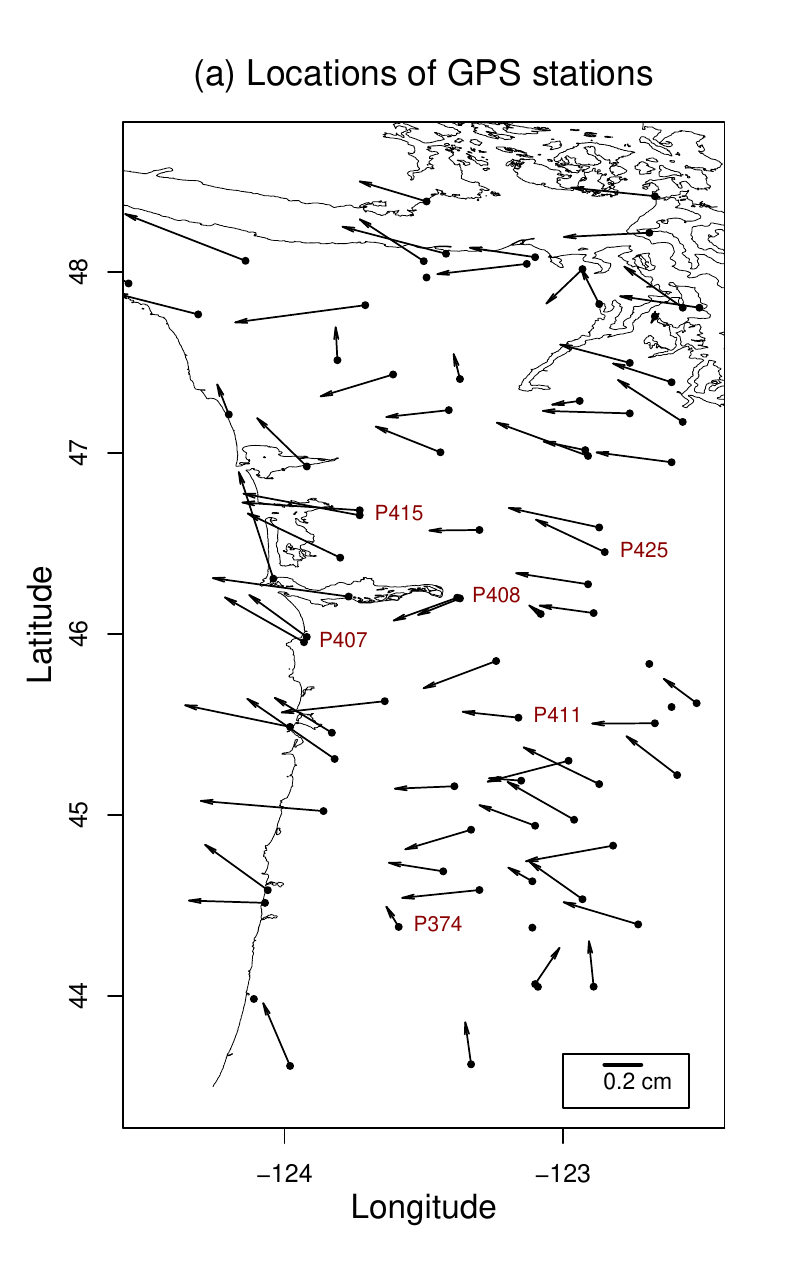}
\hspace{-.05in}
\includegraphics[scale=0.48]{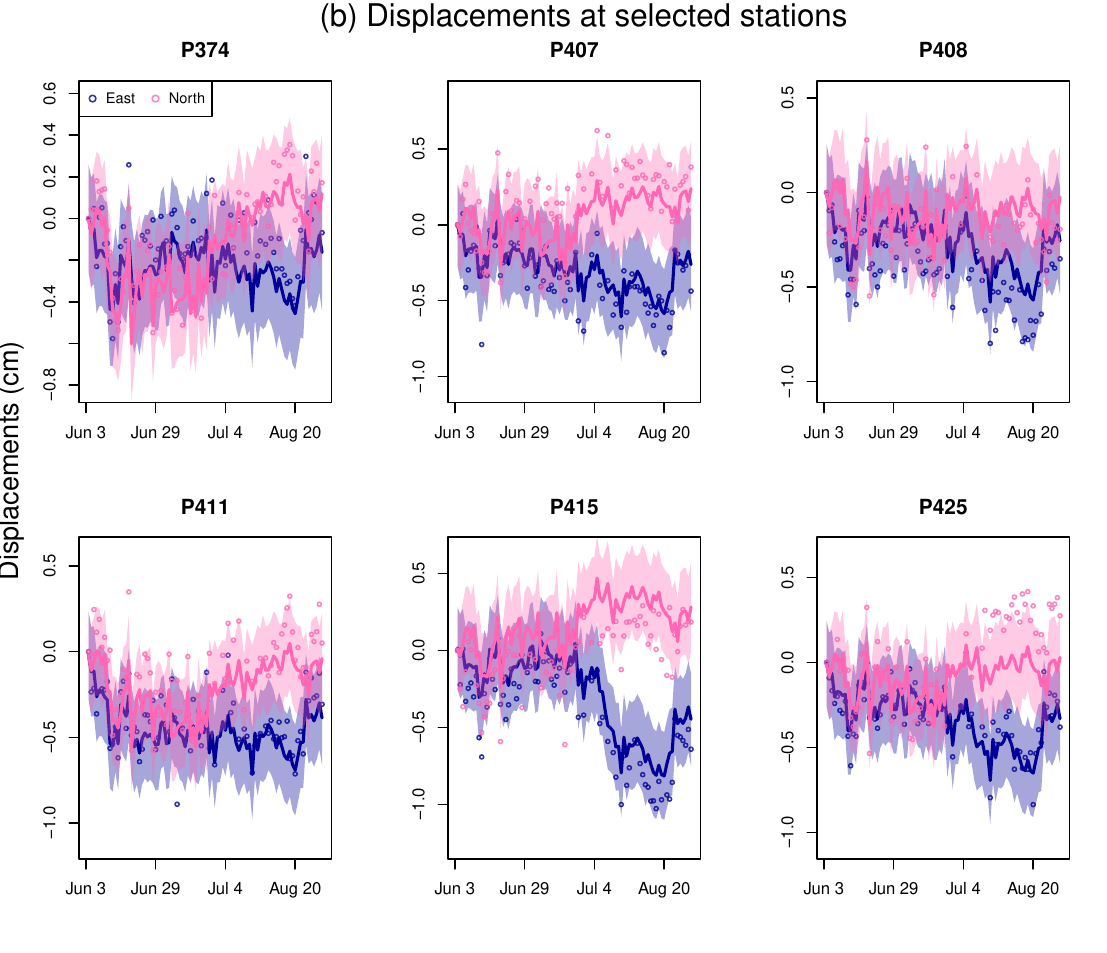}
\end{tabular}
\vspace{-.25in}
\caption{(a) 
The cumulative displacements from June 3, 2011 to August 30, 2011 in the Cascadia region are plotted by dark arrows with the unit of the measurements given in the inset. (b) The displacements by six GPS stations in East-West and North-South directions during the 2011 episodic tremor and slip event in centimeters (cm). The solid lines and corresponding shaded regions represent the predictive mean of the data and the 95\% posterior credible intervals by FMOU, respectively. 
}
\label{fig:real_data_plot}
\end{figure}

We employ GPS measurements to estimate slip rates within the Cascadia region, situated along the western edge of North America. The  Cascadia region is characterized by the subduction of the Juan de Fuca plate beneath the North American plate. This geological interaction triggers slow slip events that last for weeks to months. 
The dataset used in this study is publicly available at the Plate Boundary Observatory, which contains observations from $\tilde{k}=100$ GPS stations over $n=88$ days in 2011 \cite{bartlow2011space}. 
Each observation represents the daily average geographical position recorded in three directions: East-West, North-South and Up-Down. 
We follow \cite{bartlow2011space} to use GPS measurements in the East-West and North-South directions, 
as the vertical displacement measurements contain little information for slip estimation. Figure \ref{fig:real_data_plot}(a) shows the cumulative displacements in the Cascadia region between June 3, 2011 to August 30, 2011, while the displacements in the East-West and North-South directions of six GPS stations are shown in Figure \ref{fig:real_data_plot}(b). As identified in  \cite{demets2010geologically}, the observations encompass secular signals, such as annual and semiannual components, which are typically removed before the analysis \cite{bartlow2011space}. Details of preprocessing GPS measurements before modeling are provided in Section SM6  in the supplement materials. Apart from the GPS measurements, we utilize determinations of tectonic tremor locations from the same region and period \cite{wech2008automated}  as an independent assessment of the location of geologic slip at a given time.  

The Green's functions here are built on a mesh consisting of $k'= 1978$ triangular 
patches \cite{fukushima2005finding} located on a fault plane beneath the ground and they are calculated by assuming triangular dislocations in a homogeneous and elastic half-space \cite{thomas1993polu3d}. 
Observations from 100 GPS stations in two directions yield a 
$200 \times 1978$ Green’s function matrix $\mathbf{G}$.

Furthermore, measurements are typically influenced by assigning the GPS-derived displacements in a North American Plate fixed reference frame, and local displacements of the GPS monuments. Thus it is common to include the frame motion, a time-dependent trend of each direction shared by each GPS station \cite{bartlow2011space}. Consequently, the augmented model can be written as $ \mathbf y(t) = {\mathbf G}^{aug} {\mathbf z}^{aug}_s(t) +  \bm \epsilon(t)$, where ${\mathbf G}^{aug} =(\mathbf G, \mathbf I_{k \times 2})$ with $\mathbf I_{k \times 2} := (\mathbf I_2, \dots, \mathbf I_2)^T$, and  ${\mathbf z}^{aug}_s(t) = 
       [\mathbf z_s(t)^T,
        f_E(t),
        f_N(t)]^T$ with $f_E(t)$ and $ f_N(t)$ being the frame motion in the East-West and North-South directions, respectively, 
for $t=1,\ldots,n$. The Gaussian noise vector follows $\bm \epsilon(t) \sim N(0, \sigma_0^2 \mathbf I_k)$ with $\sigma_0^2 $ being the observed variance. 
Following \cite{bartlow2011space}, we estimate the magnitude of the geologic slips projected onto a horizontal plane at 52 degrees clockwise from North along with fault direction \cite{demets2010geologically}. 

We compare three methods. For FMOU, we use the VM method in  (\ref{equ:d_est_known_var}) for estimating the number of latent factors and 
the parameters are estimated by Algorithm \ref{algorithm:EM_FMOU_est_U}. 
We use the posterior mean of slip $\mathbf{\hat z}_s(t) $  from  (\ref{equ:posterior_dist_slips}) for estimation. Then the slip rates are calculated by 
\begin{equation}
 {\mathbf{\hat z}}_{s,rate}(t) = 
{\mathbf{\hat z}_s(t+1)-\mathbf{\hat z}}_s(t),
\label{equ:slip_rates}
\end{equation}
for $t =1, \ldots, n-1$. Following \cite{bartlow2011space}, we truncate the negative slip rates to zero, as it is deemed unlikely that the fault would slip backward in the prevailing tectonic stress state.  For comparison, we include the NIF model \cite{segall1997time} and the modified NIF model \cite{bartlow2011space}, which are summarized in Section SM4.2 
and Section SM4.3 
in the supplementary materials, respectively.   We do not include EM-VAR(1) in the real example as it requires inverting a $k'\times k'$ matrix for each time in the RTS smoother in the EM algorithm, which is too costly to compute.

\subsection{Results}
\label{sec:real_data_result}
To evaluate the model performance, we use the held-out tremor locations to validate the estimation, as tremor is well associated in both space and time with geologic slip. The tremors are monitored continuously in the Cascadia zone \cite{wech2008automated, wech2010interactive}, available from the Pacific Northwest Seismic Network catalog. We consider two metrics: the proportion of tremors identified by large slip rates with overlapping grids (the true positive rate), and the total grids containing large slip rates (the positive rate). 

\begin{figure}[t]
    \centering
\includegraphics[scale=0.6]{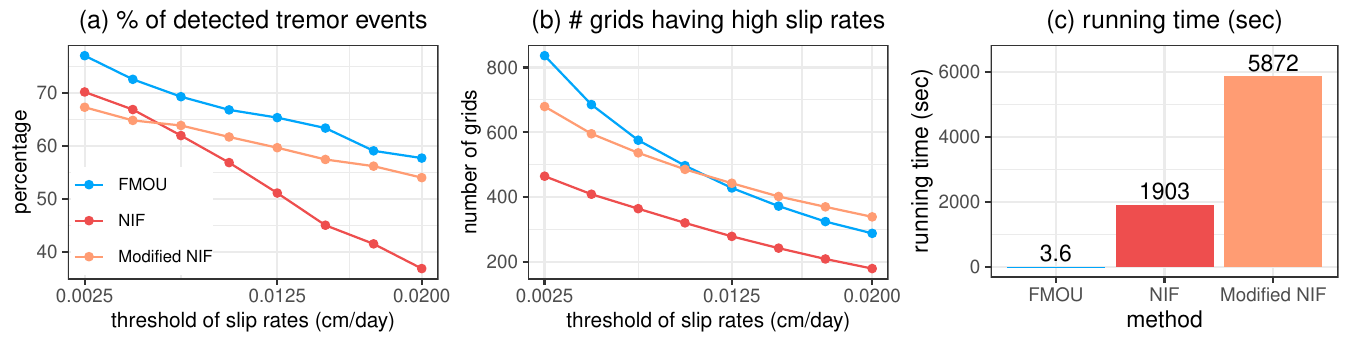}
\vspace{-.05in}
    \caption{Model performance in 2011 Cascadia data. We compare three methods: FMOU (blue), NIF (red), and modified NIF (orange). (a) Proportion of identified tremor. (b) The number of grids containing large average slip rates. (c) Running time (seconds).}
    \label{fig:real_data_alarm}
\end{figure}

Panels (a) and (b) in Figure \ref{fig:real_data_alarm} show the proportion of detected tremor events and the number of grids that contain high slip rates. 
The FMOU model detected the highest number of tremor events among all methods across all thresholds of the slip rates. A slip rate larger than 0.0125 cm/day is considered to be high. The FMOU model has a smaller number of 
spatial grids detected to have a high slip rate and it detects around  5\% more tremors than the modified NIF model, shown in panel (b) of Figure \ref{fig:real_data_alarm}. The NIF model has a substantially lower true positive rate compared to the modified NIF and FMOU, as it estimates slip rates to be negative or close to zero for more grids than the other two models. Because the noise in GPS measurement is relatively large compared to the slip-generated signal, a flexible model is preferred for capturing the heterogeneous slip changes. The FMOU model, with distinct correlation and variance parameters for each latent process, is more flexible than the NIF and modified NIF for modeling the slip propagation. 

The computational time of the three methods is given in panel (c) in Figure \ref{fig:real_data_alarm}.  Notably, the computational time of the FMOU is more than {528} times and {1631} times faster than the NIF model and the modified NIF model, respectively. The most computationally intensive step for the FMOU is estimating the number of factors, which takes {3.03} seconds. After selecting the number of latent processes, estimating all parameters and slip rates requires only {0.54} seconds. This dramatic acceleration in computation enables the use of massive datasets from large GPS networks, and  provides new opportunities to jointly estimate the hazards across larger regions, which could otherwise be prohibitive due to the large computational expense. 

\begin{figure}[t]
\centering
\begin{tabular}{c}
\includegraphics[scale=0.44]{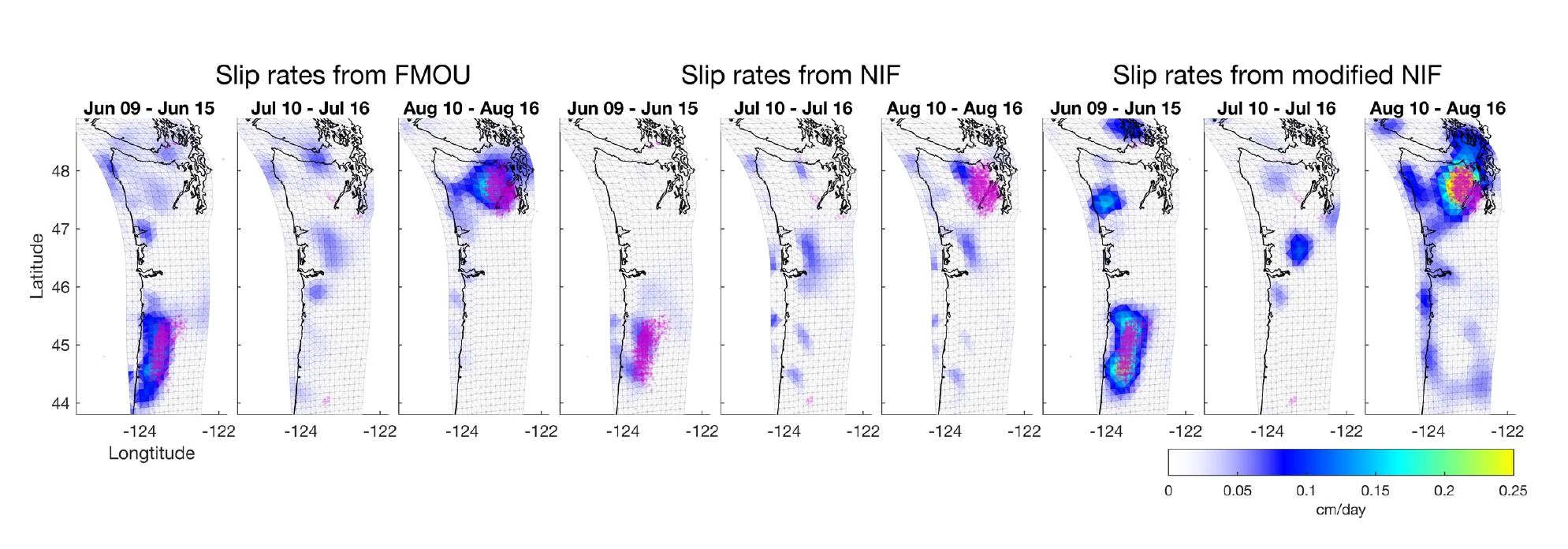}
\end{tabular}
\vspace{-.25in}
\caption{
The seven-day averages of slip rates estimated by the FMOU, NIF and modified NIF  are plotted in the left 3, middle and right 3 panels, respectively, for the Cascadia displacement data in 2011. The tremor epicenters are plotted as magenta dots. 
}
\label{fig:FMOU_slip_rate_7days}
\end{figure}

Figure \ref{fig:FMOU_slip_rate_7days} shows the seven-day average of the estimated slip rates of three periods by the FMOU, NIF and modified NIF models, 
where the solid dots represent the tremor epicenters.  Estimation of slip rates of other periods is provided in Section SM6 
in supplementary materials.  The estimates from FMOU align well with tremor dataset events,  even though the GPS measurements and tremor datasets were collected independently from distinct sources. Compared to the modified NIF, the FMOU model provides better agreement between high-slip regions and tremor epicenters during August 10–16. In contrast, the modified NIF model detects a much larger area with relatively high slip rates, particularly in the southeast region during August 10–16, resulting in a higher false positive rate. 
Both FMOU and modified NIF reveal an area of high slip rates centered around $45^{\circ}$N and $123.5^{\circ}$W in early June, and another region with high slip rates centered around $47.5^{\circ}$N and $123^{\circ}$W in early August. However, the migration of the estimated high slip rate region is less apparent in the NIF model and its modified version. The finding by the FMOU model is consistent with the previous results by \cite{bartlow2011space}, which illustrates the physical mechanism of the coincidence between tremor centers and the regions with high slip rates.

\section{Conclusion}
\label{sec:conclusion}
We proposed the FMOU approach for estimating high-dimensional dynamical systems with noisy observations. We assumed a latent factor model with an orthogonal factor load matrix, where each latent process is modeled by an OU process with distinct correlation and variance parameters estimated from  data. 
We developed a fast EM algorithm where each iteration contains closed-form expressions for parameter estimation, and it does not require inverting the covariance matrix at each time in  KF, as required in the conventional EM algorithm for  vector autoregressive models. Extensive numerical results illustrate the high efficiency and scalability of the FMOU approach compared to other methods. 

This study opens the door to various research problems in both methodology and applications. First, the OU process is not mean-squared differentiable, so it is of interest to extend the closed-form expressions for parameter estimation to differentiable Gaussian processes with Mat{\'e}rn covariance. Second, generalizing the fast algorithm to handle observations with irregular missing values is another interesting topic. Third, the FMOU implicitly induces a vector autoregressive model with a symmetric transition matrix in (\ref{eq:PDMD_2}) for the mean of the observations.  The high computational scalability and closed-form estimation of parameters make it appealing for applications such as scalable estimators for Granger causality from noisy data. Finally, for the application of slip estimation, the significantly faster approach by FMOU enables integrating geodetic data from large GPS networks, seismic data, and satellite interferograms for geological hazard quantification. 

\section*{Acknowledgement} 
We acknowledge NSF DMS-2053423 and OAC-2411043 for supporting this project. 
We thank the Editor, Associate Editor, and three anonymous referees  for their comments that substantially improved this
article.

\bibliographystyle{siamplain}
\bibliography{References_chronical_2023.bib}

\end{document}